\documentclass[12pt]{article}
\title{The extremal black holes of N=4 supergravity from so(8,2+n) nilpotent orbits}
\usepackage{mathrsfs}
\usepackage{amsmath}

\global\arraycolsep=1pt
\usepackage[colorlinks=true,backref=true,linkcolor=black,anchorcolor=black,citecolor=black,filecolor   =black,menucolor=black,pagecolor=black,urlcolor=black]{hyperref}
\usepackage[Symbol]{upgreek}
\usepackage{bbm}
\usepackage{dsfont}
\usepackage{amssymb}
\usepackage{textcomp}
\usepackage{wasysym}

\newcommand{\ba}{/ \hspace{-1.2ex}} 
\newcommand{\baa}{/ \hspace{-1.4ex}}
\newcommand{\baaa}{\, / \hspace{-1.6ex}}

\def\bahQ{ \hspace{.3mm} /  \hspace{-1.7mm}   \hat{Q}  \hspace{0.2mm}}

\def\baahP{\hspace{0.7mm}  /  \hspace{-1.9mm} \hat{P}  \hspace{0.3mm}}

\def\bahp{ /  \hspace{-1.7mm}   \hat{p}  \hspace{0.5mm}}
\def\bahq{ /  \hspace{-1.5mm}   \hat{q}  \hspace{0.5mm}}

\def\baaaP{ \baaa \hspace{.2mm}P}
\def\baaahP{ \baaa \hspace{.2mm} \hat{P}}
\def\baaU{ \hspace{.3mm} \baa U}

\newcommand{\Scal}[1]{\Bigl ({#1} \Bigr )}
\newcommand{\scal}[1]{\bigl ({#1} \bigr )}
\def\bea{\begin{eqnarray}}
\def\eea{\end{eqnarray}}
\def\be{\begin{equation}}
\def\ee{\end{equation}}
\newcommand{\CR}{\nonumber \\*}

\newcommand{\trace}{\hbox {Tr}~}

\DeclareMathAlphabet{\mathpzc}{OT1}{pzc}{m}{it}

\DeclareMathOperator{\ad}{ad}

\newcommand{\ord}[1]{{\scriptscriptstyle (#1)}}

\def\C{\mathscr{C}}

\usepackage[Symbol]{upgreek}
\usepackage{caption}
\usepackage{bbm}
\usepackage{graphicx}

\def\un{{\mathpzc{1}}}
\def\deux{{\mathpzc{2}}}

\newcommand{\sfrac}[2]{{\scriptstyle \frac{#1}{#2}}}

\def\DJo{$\;$\kern-.4em \hbox{D\kern-.8em\raise.15ex\hbox{--}\kern.35em okovi\'c}}

\def\DEVIII#1#2#3#4#5#6#7#8{{\tiny $ { \left[ \begin{array}{ccccccc}  & & \mathfrak{#2} \hspace{-0.7mm}&&&& \vspace{ -1.5mm} \\ \mathfrak{#1}\hspace{-0.7mm} &  \mathfrak{#3} \hspace{-0.7mm}& \mathfrak{#4} \hspace{-0.7mm} & \mathfrak{#5}\hspace{-0.7mm}&\mathfrak{#6}\hspace{-0.7mm}&\mathfrak{#7}\hspace{-0.7mm}&\mathfrak{#8} \end{array}\right] }$}}
\def\DSOXVI#1#2#3#4#5#6#7#8{{\tiny $ {  \vspace{-2mm} \left[ \begin{array}{ccccccccc}  && \mathfrak{#8} \hspace{-0.7mm}&&&&&& \vspace{ -1.5mm} \\ \cdot \hspace{-0.5mm}& \mathfrak{#7}\hspace{-0.7mm} &\mathfrak{#6}\hspace{-0.7mm} &  \mathfrak{#5} \hspace{-0.7mm}& \mathfrak{#4} \hspace{-0.7mm} & \mathfrak{#3}\hspace{-0.7mm}&\mathfrak{#2}\hspace{-0.7mm}&\mathfrak{#1} \end{array}\right] }$}}

\def\ie{{\it i.e.}\ }
\def\eg{{\it e.g.}\ }

\def\nn{\nonumber}

\def\N{\mathcal{N}}

\def\ft#1#2{\tfrac{#1}{#2}}

\def\C{{\mathscr{C}}}
\def\V{{\mathcal{V}}}
\def\w{{\scriptstyle W}}

\def\G{{\mathfrak{G}}}
\def\H{{\mathfrak{H}}}

\def\m{{\mathpzc{m}}}
\def\n{{\mathpzc{n}}}
\def\p{{\mathpzc{p}}}

\def\a{{\mathpzc{a}}}
\def\b{{\mathpzc{b}}}
\def\zero{{\mathpzc{0}}}
\def\un{{\mathpzc{1}}}
\def\deux{{\mathpzc{2}}}

\def\Ic{{I\hspace{-0.6mm}c}}

\def\DJo{$\;$\kern-.4em \hbox{D\kern-.8em\raise.15ex\hbox{--}\kern.35em okovi\'c}}

\newcommand{\eprint}[1]{{\href{http://arxiv.org/abs/#1}{\texttt{[#1}]}}}
\newcommand{\eprintN}[1]{{\href{http://arxiv.org/abs/#1}{\texttt{#1 [hep-th]}}}}

\def\e{\boldsymbol{e}}

\def\h{\boldsymbol{h}}

\def\z{\boldsymbol{z}}

\def\g{\mathfrak{g}}
\def\G{\mathfrak{G}}
\def\h{\mathfrak{h}}
\def\H{\mathfrak{H}}

\def\ft#1#2{${\textstyle{{\scriptstyle #1}\over {\scriptstyle #2}}}$}

\def\w{{\scriptstyle W}}
\def\Ic{{I\hspace{-0.6mm}c}}

\def\gl{\mathfrak{gl}}
\def\sl{\mathfrak{sl}}
\def\so{\mathfrak{so}}
\def\su{\mathfrak{su}}

\def\e{\mathfrak{e}}

\def\Ha{\mathcal{H}}
\def\Ka{\mathcal{K}}
\def\nn{\nonumber}

\def\N{\mathcal{N}}

\begin{document}
\allowdisplaybreaks[1]
\renewcommand{\thefootnote}{\fnsymbol{footnote}}
\numberwithin{equation}{section}
\def\corr{$\spadesuit \, $}
\def\trefle{$ \, $}
\def\kscorr{$\diamondsuit \, $}
\begin{titlepage}
\begin{flushright}
\
\vskip -2.5cm
{\small AEI-2009-057}\\
\vskip 1cm
\end{flushright}
\begin{center}
{\Large \bf
The extremal black holes of $\N=4$ supergravity \\* \vskip 2mm from $\so(8,2+n)$ 
nilpotent orbits}
\\
\lineskip .75em
\vskip 3em
\normalsize
{\large  Guillaume Bossard\footnote{email address: bossard@aei.mpg.de}}\\
\vskip 1 em
$^{\ast}${\it AEI, Max-Planck-Institut f\"{u}r Gravitationsphysik\\
Am M\"{u}hlenberg 1, D-14476 Potsdam, Germany}
\\

\vskip 1 em
\end{center}
\begin{abstract}
{\footnotesize
We consider the stationary solutions of $\N=4$ supergravity coupled to $n$ vector multiplets that define linear superpositions of non-interacting extremal black holes. The most general solutions of this type are derived from the graded decompositions of $\so(8,2+n)$ associated to its nilpotent orbits.  We illustrate the formalism by giving explicitly asymptotically Minkowski non-BPS solutions of the most exotic class depending on $6+n$ harmonic functions.

}
\end{abstract}

\end{titlepage}
\renewcommand{\thefootnote}{\arabic{footnote}}
\setcounter{footnote}{0}


\section{Introduction}
When considering BPS solutions within supergravity theories, one can solve the Einstein equations by considering the much simpler first order equations defining the supersymmetry variations of the fermions. For example, the most general BPS asymptotically Minkowski black holes of pure $\N=4$ supergravity depending on four harmonic functions have been derived in \cite{Ortin}. Nonetheless, the equations of motions of $\N=4$ supergravity coupled to $n$ vector multiplets are not very different from the more general ones of gravity coupled to $n+m$ abelian vector fields and scalar fields parametrising a symmetric space of the form 
\be SL(2,\mathds{R})/ SO(2) \cup SO(m,n)/ (SO(m)\times SO(n)) \label{MN} \ee
 which only define the bosonic sector of supersymmetric theories for $m=2$ and $6$ (which are then $\N=2$ and $\N=4$, respectively). One would thus expect to be able to derive such solutions of Papapetrou--Majumdar type \cite{Papapetrou,Majumdar} systematically, without referring to supersymmetry.

Stationary solutions of $\N=4$ supergravity coupled to $n$ vector multiplets satisfy the equations of motion of a non-linear sigma model defined over the pseudo-Riemanniann symmetric space $SO(8,2+n)/ ( SO(6,2)\times SO(2,n))$ coupled to Euclidean gravity in three dimensions. Within this formulation, the construction of multi-black hole solutions depending on arbitrary harmonic functions amount to resolving an algebraic equation (\ref{tricom}) \cite{Clement}. It has been explained in a recent publication \cite{PapaMajumdar}Ê  that the general solutions of this equation can be derived from the graded decompositions of the Lie algebra $\so(8,2+n)$ associated to its nilpotent orbits \cite{PapaMajumdar} (or more generally of the simple Lie algebra $\g$ for any non-linear sigma model over a symmetric space  $\G / \H^*$). Moreover, referring to the general classification of \cite{nous}, one can distinguish from those solutions which are the ones that define regular space-time in which all the singularities are covered by an horizon.

One motivation for considering the particular example of $\N=4$ supergravity comes from the recent discovery of non-BPS extremal solutions within the $STU$ model \cite{DallAgatta,Hotta,Gimon} (\ie for $m=n=2$ within (\ref{MN})). Such extremal solutions can be derived from a `fake superpotential' \cite{STU} within the formalism of the attractor mechanism \cite{attractors,attractors1}. A systematic way of deriving the attractor superpotential from the nilpotent orbit of the Noether charge may shed some light on the counting of non-BPS black holes microstates \cite{Dabholkar}.

From another point of view, $\N=4$ supergravity theories provide a large class of theories that can be studied in the framework of \cite{PapaMajumdar}, and which are simple enough to allow for an explicit computation of the solutions. For instance, the general method defined in \cite{PapaMajumdar} permits to derive in a straightforward way the solutions of the three-dimensional non-linear sigma model in the symmetric gauge. Nevertheless, in order to read of the explicit solution in term of the four-dimensional fields, and thus to extract the physical quantities such as the horizon area of the black holes, one must then rotate the coset representative into a specific parabolic gauge. This last step can be technically difficult, as for instance in the case of maximal supergravity for which one must consider the multiplication of elements of $E_{8(8)}$ which smallest irreducible representation is $248$-dimensional. As we are going to see, such computation can be carried out much more easily in the case of orthogonal groups. 

The paper starts by a brief revue of the method derived in \cite{PapaMajumdar}. We then display in detail the different classes of solutions of Papapetrou--Majumdar type of $\N=4$ supergravity. They come into three classes, the first one being the known linear superpositions of \ft14 BPS black holes preserving four identical supersymmetry charges. The second corresponds to linear superpositions of non-BPS black holes, which central charges vanish at the horizon. The latter can be understood from the former within $\N=4$ supergravity coupled to $6$ vector multiplets, by simply permuting the vector fields belonging to the gravity multiplets and those belonging to the vector multiplets. More generally, they are \ft12 BPS within an $\N=2$ supergravity theory which bosonic sector defines a consistent truncation of the $\N=4$ theory, such that the $\N=2$ graviphoton lies inside a vector multiplet of the latter. The last class corresponds to linear superpositions of non-BPS black holes which central charges are non-zero at the horizon. These more exotic solutions are not BPS solutions of an appropriated $\N=2$ truncation of the theory and involve the whole field content of the latter \cite{PapaMajumdar}. We will give explicitly a large class of such solutions depending on $6+n$ harmonic functions in section \ref{solution}.

It is argued in \cite{PapaMajumdar} that the most general regular Papapetrou--Majumdar type solutions can be extracted from the five-graded decomposition of $\so(8,2+n)$ associated to the dimensional reduction from four to three dimensions (\ref{fivegrade}). We prove in the last section that all the solutions involving higher order nilpotent orbits do indeed carry naked singularities. We also provide some strong evidence that the result extends to maximal supergravity as well.

\section{Extremal solutions and nilpotent orbits}
The bosonic field content of $\N=4$ supergravity is given by the gravity multiplet, that is the metric, six abelian vector fields, and the axion / dilaton scalar fields parametrising the symmetric space $SL(2,\mathds{R})/ SO(2) $, and by $n$ vector multiplets, containing $n$ abelian vector fields and scalar fields parametrising the symmetric space $SO(6,n) / ((SO(6) \times SO(n))$.  The $6+n$ vector fields transform in the vector representation of the isometry group $SO(6,n)$, and $SL(2,\mathds{R})$ mixes the `electric' and the `magnetic components'. The stationary solutions of the theory satisfy the equations of motion of a non-linear sigma model over $SO(8,2+n) / ( SO(6,2) \times SO(2,n))$ coupled to Euclidean gravity in three dimensions.  For a coset representative $\V$ in  $SO(8,2+n) / ( SO(6,2) \times SO(2,n))$, one decomposes the Maurer--Cartan form $\V^{-1} d \V$ into its coset and its $\so(6,2) \oplus \so(2,n)$ components,
\be
\V^{-1} d\V = Q + P \hspace{10mm} \begin{array}{l} Q\equiv Q_\mu dx^\mu\in \so(6,2) \oplus \so(2,n)  
 \\ P\equiv P_\mu dx^\mu\in \so(8,2+n) \ominus \so(6,2) \oplus \so(2,n) \end{array}
\ee
and the equations of motion read
\be
d   \star \V  P  \V^{-1}Ê = 0 \hspace{20mm} R_{\mu\nu}  = \trace P_\mu P_\nu \label{EinsteinE}
\ee
where $\star$ is the Hodge star operator associated to the three-dimensional 
Riemannian metric $g$.

In order to exhibit the field content of the four-dimensional theory  it is convenient to consider the coset representative $\V$ in the spinor representation of $Spin(8,2+n)$. One defines then the Clifford algebra of $Spin(8,2+n)$ as the tensor product of the Clifford algebra of $Spin(2,2) \cong SL(2,\mathds{R}) \times SL(2,\mathds{R}) $ and the Clifford algebra of $Spin(6,n)$, such that $SL(2,\mathds{R}) \times SL(2,\mathds{R}) \times Spin(6,n)$ is the product of the Ehlers group and the duality symmetry group of the four-dimensional theory. This way one writes the coset representative $\V$ as a four by four matrix valued in the Clifford algebra of $Spin(6,n)$,
\be \V =  \left(\begin{array}{cccc} \hspace{2mm}H^{\frac{1}{2}}  \hspace{2mm}& \hspace{2mm} H^{-\frac{1}{2}} \scal{ B - \sfrac{1}{2} [  \baaU , \baaa A] }\hspace{2mm}& \hspace{2mm}  \bar H^{\frac{1}{2}}  \baaU  \hspace{2mm}& \hspace{2mm} \bar H^{-\frac{1}{2}} \scal{ \baaa A + \bar B  \baaU }  \hspace{2mm}  \\
 \hspace{2mm} 0 \hspace{2mm} &\hspace{2mm} H^{-\frac{1}{2} } \hspace{2mm} & \hspace{2mm}0\hspace{2mm} & \hspace{2mm} 0 \hspace{2mm} \\  0  & - H^{- \frac{1}{2} } \baaa A  & \bar H^{\frac{1}{2} }  & \bar H^{-\frac{1}{2} } \bar B  \\ 0  &  H^{- \frac{1}{2} }  \baaU  & 0  & \bar H^{-\frac{1}{2} } \end{array} \right)  v \label{KKa} \ee
 where $v$ is the coset representative of the scalars in $SO(6,n) / ( SO(6) \times SO(n))$ in the spinor representation of $Spin(6,n)$, and $\bar H$ and $\bar B$ are the dilaton and the axion fields in the parabolic gauge of $SL(2,\mathds{R})/ SO(2)$. Note that we identify the identity $\mathds{1}$ of the Clifford algebra associated to $Spin(6,n)$ with the real unit $1$ in the formulas. The electric and the magnetic components $\baa U$ and $\baaa A$ of the vector fields are contracted with the $Spin(6,n)$ gamma matrices. After dualization of $B$ according to its equation of motion,
 \be d \hat{B} = -  H^{-2} \star  \Scal{ d B + \frac{1}{2}Ê \{ \baa U ,   d  \baaa A \} -  \frac{1}{2}Ê\{Ê \baaa A , d  \baa U \}  } 
\ee 
one recovers the vector field $\hat{B}_\mu $ defining altogether with $H$ the metric through the Kaluza--Klein ansatz    
 \be ds^2 = - H \scal{ dt + \hat{B}_\mu dx^\mu }^2 + H^{-1} g_{\mu\nu} dx^\mu dx^\nu \ee
And dualizing as well the $SO(6,n)$ vector $\baaa A$ according to its equation of motion, 
\be   \scal{  d  \baaa \hat{A} + \baa U  d \hat{B}}= - ( H \bar H )^{-1} \star \scal{d \baaa A  +  \bar B d \baa U }  \ee
one recovers the vector fields $\baaa \hat{A}_\mu$ defining altogether with $\baaU$ the $SO(6,n)$ vector $\mathcal{A}^I$ of abelian vector fields, through the Kaluza--Klein ansatz
\be \sqrt{ 8\pi G} \mathcal{ A}^I  = \{ \Gamma^I , \baaU  \} \scal{Êdt + \hat{B}_\mu dx^\mu}  + \{Ê\Gamma^I , \baaa \hat{A}_\mu \} dx^\mu \ee

The spherically symmetric black holes (including the asymptotically Taub--NUT ones) are entirely 
characterised by their $SO(8,2+n)$-Noether charge
\be 
\mathcal{Q} \equiv  \frac{1}{4\pi} \int_{\partial V}  \star \V P \V^{-1}  
\ee
and the asymptotic value of the scalars fields $ \V_\zero \in SL(2,\mathds{R}) \times  SO(6,n)$ 
at spatial infinity. Actually, it will be more convenient to characterise 
the solutions in term of a modified conserved charge $\C$ obtained by rotating 
$\mathcal{Q}$ back into the coset
\be 
\C \equiv   {\V_\zero}^{-1}    \mathcal{Q}  \, \V_\zero  
\in \so(8,2+n)  \ominus \scal{Ê\so(6,2) \oplus \so(2,n) }
\ee
which we will call the `Noether charge' for simplicity (this designation being 
unambiguous since we will never refer to $\mathcal{Q}$ itself). It has the following form 
\be 
\C= \left( \begin{array}{cccc} \,  M  + \ba \sigma \, & \,   N \, & \, -  
\, \baa Q  + \ba q \, & \, \ba p -   \baaaP \, \\
\,  N \, & -  M  + \ba \sigma & - \ba p -  \baaaP &  \, \baa Q + \ba q \\  \, \baa Q  + \ba q &   
- \ba p + \baaaP   &  \Sigma  + \ba \sigma &   \Xi \\
   \baaaP  + \ba p & -  \, \baa Q  + \ba q &   \Xi & -  \Sigma  + \ba \sigma \end{array} \right) \label{Noether} 
\ee 
where $M$ and $N$ are the mass and the NUT charge, $\baa Q$ and $\baaaP$ are the $SO(6)$ vectors of electric and magnetic components of the central charges contracted with the $Spin(6,n)$ gamma matrices, $\ba q$ and $\ba p$ are the $SO(n)$ vectors of electric and magnetic charges of the vector fields of the vector multiplets generalising the central charges, contracted with the $Spin(6,n)$ gamma matrices as well. $\Sigma$ and $\Xi$ are the dilaton and the axion charge, respectively, and $\ba \sigma$ is the $\so(6,n) \ominus \scal{Ê\so(6) \oplus \so(n)}$ charge associated to the coset scalars of $SO(6,n) / ( SO(6) \times SO(n))$. Note that we have rescaled all the electro-magnetic charges by a factor of $\sqrt{2}$ with respect with the usual conventions \cite{Ortin} in the sake of simplicity of the formulas. The reader must also keep in mind that $\baa Q$, $\baaaP$, $\ba q$ and $\ba p$ are the charges in the vector representations of $SO(2) \times SO(6) \times SO(n)$ that generalise the central charges, and not the electromagnetic charges transforming with respect with $SL(2,\mathds{R}) \times SO(6,n)$.

As stated in \cite{nous}, regular black holes admit a Noether charge $\C$ which satisfy the characteristic equation 
\be \C^3 = c^2 \C \label{cubic} \ee
where $c$ defines the normalised trace of $\C^2$ 
\be c^2 = M^2 + N^2 - 2  Q^2 - 2 P^2 - 2 q^2 - 2 p^2 + \Sigma^2 + \Xi^2 \ee
which is proportional to the product of the horizon area $ A_{\scriptscriptstyle \mathscr{H}}$ and the surface gravity  $\upkappa$ for regular spherically symmetric black holes \cite{PapaMajumdar},
\be A_{\scriptscriptstyle \mathscr{H}}  \upkappa = 4\pi \, c \label{extremal} \ee
The cubic equation (\ref{cubic}) determines the scalar charges $\Sigma$, $\Xi$ and $\ba \sigma$ in function of the others, although they are irrational functions of the mass, NUT and electromagnetic charges in general.  For extremal black holes, it follows from (\ref{cubic}) and (\ref{extremal})  that the Noether charge is then nilpotent
\be \C^3 = 0 \label{tripot} \ee
and one can then write done the expression of the scalar charges in function of the others in closed form. 

The complex $O(10+n,\mathds{C})$ orbit of a generic solution of equation (\ref{tripot}) is dense in the set of solutions of this equation in $\so(10+n,\mathds{C})$ \cite{coadjoint}. To any representative ${\bf E}$ of a general nilpotent orbit of $\so(10+n,\mathds{C})$, one can associate a corresponding $\sl_2(\mathds{C})$ triplet $({\bf H},{\bf E},{\bf F})$ such that  \cite{coadjoint}
\be [ {\bf H} ,{\bf  E} ] = 2 {\bf E} \hspace{10mm} [{\bf H},{\bf F}] = -2 {\bf F} \hspace{10mm} [{\bf E},{\bf F}] = {\bf H} \label{triplet} \ee
and such that ${\bf H}$ defines a graded decomposition of $\so(10+n,\mathds{C})$ which characterises uniquely the orbit. When the $O(10+n,\mathds{C})$ orbit admits a non-trivial intersection with $\so(8,2+n)$, the triplet can be chosen to define an $\sl_2$ triplet of $\so(8,2+n)$ and ${\bf H}$ defines a unique graded decomposition of $\so(8,2+n)$. The graded decomposition associated to equation (\ref{tripot}) is the one associated to the dimensional reduction from four to three dimensions,
\be \so(8,2+n) \cong {\bf 1}^\ord{-4} \oplus \scal{Ê{\bf 2} \otimes ({\bf 6+n})}^\ord{-2} \oplus  \scal{\gl_1 \oplus Ê\sl_2 \oplus \so(6,n)}^\ord{0}  \oplus \scal{Ê{\bf 2} \otimes ({\bf 6+n})}^\ord{2} \oplus{\bf 1}^\ord{4} \label{fivegrade}\ee
Representatives of the orbits are generic elements of the grade two component (\ref{triplet}) that define $\sl_2$ doublets of orthogonal non-null vectors of $SO(6,n)$. For $n> 2$ there are three real orbits of $SO(8,2+n)$ associated to this graded decomposition. They correspond to $\sl_2$ doublets of orthogonal non-null vectors of $SO(6,n)$ in $ \scal{Ê{\bf 2} \otimes ({\bf 6+n})}^\ord{2}$ which are either both time-like (\ie of isotropy subgroup $SO(4,n) \subset SO(6,n)$), both space-like (\ie of isotropy subgroup $SO(6,n-2) \subset SO(6,n)$), or of mixed type (\ie of isotropy subgroup $SO(5,n-1) \subset SO(6,n)$). They are are commonly labelled as $(+-+)^2$, $(-+-)^2$ and $(+-+)(-+-)$, respectively \cite{coadjoint}.\footnote{For $n=2$ the real orbit of $O(8,2+n)$ associated to a doublet of space-like vectors decomposes into two orbits of the connected component $SO_\zero(8,2+n)$. For $n=1$ the latter orbit does not exist, and the one associated to a doublet of vector of mixed type decomposes into two distinct orbits of $SO_\zero(8,2+n)$. For $n=0$ there is one single orbit.}

Interestingly, each real orbit of $SO(8,2+n)$ can be associated to one single $SO(6,2) \times SO(2,n)$ orbits of regular extremal black holes \cite{nous}. In order to determine the corresponding non-trivial intersections of $  \scal{Ê{\bf 2} \otimes ({\bf 6+n})}^\ord{2} $ with $\so(8,2+n) \ominus \scal{Ê\so(6,2) \oplus \so(2,n)} $ in which the corresponding Noether charge $\C$ lies, one identifies a triplet such that both ${\bf E}$ and ${\bf F}$ lie in the coset component, and such that ${\bf H}$ lies in $Ê\so(6,2) \oplus \so(2,n)$.   As we are going to see, the orbit $(+-+)^2$ correspond to \ft14 BPS solutions, the orbit $(-+-)^2$ to non-BPS solutions for which the central charges vanish at the horizon, and the orbit $(+-+)(-+-)$ to non-BPS solutions for which the central charges do not vanish at the horizon.

We consider an Ansatz of the form 
\be 
\V =   \V_\zero \, \exp\left(- \sum_\n \Ha^\n \C_\n \right) \label{Ansatz} 
\ee
for some functions $\Ha^\n$ and Lie algebra elements $\C_\n$ all lying in the intersection of 
$  \scal{Ê{\bf 2} \otimes ({\bf 6+n})}^\ord{2} $ with $\so(8,2+n) \ominus \scal{Ê\so(6,2) \oplus \so(2,n)} $. Then, it follows from the grading (\ref{fivegrade}) that
\be [Ê\C_\m , [ \C_\n , \C_\p ] ] = 0  \hspace{20mm}Ê \trace  \C_\m \C_\n = 0  \label{tricom}\ee
such that the equations of motions reduce to the linear equations \cite{Clement}
\be R_{\mu\nu}   = 0 \hspace{20mm}  d \star d \Ha^\n = 0 \ee
One has then general Papapetrou--Majumdar solutions with $g_{\mu\nu}$ being the flat Euclidean metric and $\Ha^\n$ arbitrary asymptotically flat harmonic function of $\mathds{R}^3$. It has been argued in \cite{PapaMajumdar} that the most general solutions of this type (\ref{Ansatz},\ref{tricom}) for which all singularities are covered by an horizon are the ones associated to the three orbits  $(+-+)^2$, $(-+-)^2$ and $(+-+)(-+-)$. This proposition will be proven explicitly in the last section. We will now discuss the various solutions associated to these three orbits.

\section{BPS black holes}
We will consider first the BPS multi-black hole solutions. In this case it is well known that BPS black holes preserving four identical supersymmetry charges do not interact, such that the corresponding linear superpositions define well behaved Papapetrou--Majumdar solutions.  

For BPS solutions, it is more convenient to consider $SO(2,6)$ as the quotient of the R-symmetry group $Spin^*(8)$ by the $\mathds{Z}_{2}$ kernel of its chiral spinor representation,\footnote{We recall that $SO^*(8)$ and $SO(2,6)$ are inequivalent $\mathds{Z}_2$ quotient of $Spin^*(8)\cong Spin(2,6)$ which are related by triality.}  and to combine the charges into complex combinations 
\be \w \equiv M + i N \hspace{7mm}  Z_{ij} \equiv \frac{1}{\sqrt{2}}  [ÊC\,  \baa Q]_{ij_+} +  \frac{i}{\sqrt{2}}  [ÊC\,  \baaaP]_{ij_+}  \hspace{7mm} z^A \equiv q^A + i p^A \hspace{7mm} \varsigma \equiv \Sigma + i \Xi \ee
where we use the homomorphism $Spin(6) \cong SU(4)$ to write down the central charges as an antisymmetric tensor of $SU(4)$. We define similarly the complex selfdual tensor $\Sigma_{ij_+}^A$ from $\ba \sigma$. Using a fermionic harmonic oscillator basis for $\so^*(8)$, one can write down the Noether charge $\C$ as an $SO(2,n)$ vector of Majorana--Weyl spinors   $|\C\rangle$ of $Spin^*(8)$ \cite{nous},
\be |\C\rangle = \left(\begin{array}{ccccccc} \frac{1}{2}  \Bigl(  &\w + \bar \varsigma  &+& \scal{  Z_{ij} + \frac{1}{2}  \varepsilon_{ijkl} Z^{kl} }  Ê a^i a^j &+& \frac{1}{24} \varepsilon_{ijkl}Ê ( \varsigma + \bar \w ) a^i a^j a^k a^l &\Bigr) | 0 \rangle \\* 
\frac{i}{2}  \Bigl( & \w - \bar \varsigma  &+& \scal{  Z_{ij} - \frac{1}{2}  \varepsilon_{ijkl} Z^{kl} }  Ê a^i a^j &+ &\frac{1}{24} \varepsilon_{ijkl}Ê ( \varsigma - \bar \w ) a^i a^j a^k a^l &\Bigr) | 0 \rangle \\*
 \Bigl( & z^A & + &\Sigma_{ij+}^A  Ê a^i a^j &+ &\frac{1}{24} \varepsilon_{ijkl}Ê\bar z^A  a^i a^j a^k a^l &\Bigr) | 0 \rangle
 \end{array} \right)\ee
 which we will write
 \be |\C\rangle = \left(\begin{array}{ccccccc}  \bigl(  &\w   &+&  Z_{ij}  Ê a^i a^j &+& \frac{1}{24} \varepsilon_{ijkl}Ê  \varsigma \,  a^i a^j a^k a^l &\bigr) | 0 \rangle \\* 
 \bigl( & z^A & + &\Sigma_{ij+}^A  Ê a^i a^j &+ &\frac{1}{24} \varepsilon_{ijkl}Ê\bar z^A  a^i a^j a^k a^l &\bigr) | 0 \rangle
 \end{array} \right)\ee
for simplicity. 
\subsection{\ft14 BPS solutions}
Four real Killing spinors can be chosen in an appropriated basis to satisfy 
\be \epsilon^1_\alpha + \varepsilon_{\alpha\beta} \epsilon_2^\beta = 0 \hspace{10mm} \epsilon^3_\alpha = \epsilon^4_\alpha = 0 \ee
such that the `Dirac equation' defining the BPS condition \cite{nous} 
\be \label{Dirac} 
\scal{Ê\epsilon_\alpha^i a_i + 
\varepsilon_{\alpha\beta} \epsilon^\beta_i a^i } | \C \rangle = 0 
\ee
reads
\be   \scal{Êa_1 - a^2 }  | \C \rangle =  \scal{Êa^1 + a_2 }  | \C \rangle = 0  \label{Diracquart} \ee
The general solution of which is
\be |\C\rangle = \Scal{Ê\, ( 1 + a^1 a^2 ) \scal{Ê\w +  z  \, a^3 a^4 } |0\rangle \, ,\,  ( 1 + a^1 a^2 ) \scal{Êz^A +  \bar z^A \,  a^3 a^4 } |0\rangle \, } \label{quartBPS} \ee
From the point of view of the associated nilpotent orbit, $|\C\rangle$ is defined equivalently from the $\so^*(8)$ generator 
\be {\bf H}_\frac{1}{4} \equiv 2 a^1 a^2 - 2 a_1 a_2 \ee
by the equation\footnote{This is easily seen to be equivalent to (\ref{Diracquart}) by noting that $a^1 a^2 - a_1 a_2 = 1 - a^1 ( a_1 - a^2 ) - a_1 ( a^1 + a_2 ) $.}
\be {\bf H}_\frac{1}{4}  \, | \C \rangle = 2 | \C \rangle \ee
The generator ${\bf H}_\frac{1}{4}$ decomposes  $\so^*(8)$ as
\be \so^*(8) \cong {\bf 1}^\ord{-4} \oplus ( {\bf 2} \otimes {\bf 4})^\ord{-2} \oplus \scal{Ê \gl_1 \oplus \sl_2 \oplus \so(4) }^\ord{0} \oplus ( {\bf 2} \otimes {\bf 4})^\ord{2} \oplus {\bf 1}^\ord{4} \ee
and decomposes as well the coset component of $\so(8,2+n)$,
\be \so(8,2+n) \ominus \scal{Ê\so^*(8) \oplus \so(2,n)}  \cong \scal{ {\bf 2} \otimes ({\bf 2+n})}^\ord{-2} \oplus \scal{ {\bf 4} \otimes ({\bf 2+n})}^\ord{0} \oplus \scal{ {\bf 2} \otimes ({\bf 2+n})}^\ord{2} \ee
The general \ft14 solutions are thus defined by choosing $4+2n$ harmonic functions with associated charge matrix lying in the $ \scal{ {\bf 2} \otimes ({\bf 2+n})}^\ord{2}$ component, that are of the form (\ref{quartBPS}). 

The variety of charges lying in  $ \scal{ {\bf 2} \otimes ({\bf 2+n})}^\ord{2}$ associated to regular black holes is a stratified space which can be embedded into the stratified space of charges defining regular black holes through a filtration preserving homeomorphism. The $\mathds{R}_+^* \times SL(2,\mathds{R}) \times SO(2,n)$ orbit of a generic regular charge of $ \scal{ {\bf 2} \otimes ({\bf 2+n})}^\ord{2}$ is dense in the subset of $ \scal{ {\bf 2} \otimes ({\bf 2+n})}^\ord{2}$ associated to regular black holes. We are now going to describe the various orbits of $\mathds{R}_+^* \times SL(2,\mathds{R}) \times SO(2,n)$ of charges preserving four identical supersymmetry generators, and their embedding inside the corresponding $SO(6,2) \times SO(2,n)$ orbits of charges associated to regular extremal spherically symmetric black holes (which are displayed in \cite{nous}). 

The \ft14 BPS  multi-black holes solutions include generic \ft14 BPS black holes, whose charges satisfy\begin{gather}
 |\w|^2 > |z|^2 \hspace{20mm} 2 |\w|^2 > z^A \bar z_A + \sqrt{ ( z^A \bar z_A )^2  - | z^A z_A|^2 } \CR
 \scal{Ê|\w|^2 + |z|^2 - z^A \bar z_A }^2 >   \bigl|  2 \w \bar z - z^A z_A \bigr|^2   
 \end{gather}
where the last condition is the positivity of the $SL(2,\mathds{R}) \times SO(6,n)$ quartic invariant
\be \lozenge(\w^{-\frac{1}{2}} Z_{ij} , \w^{-\frac{1}{2}} z^A ) \equiv | \w|^{-2}Ê  \scal{Ê2 Z_{ij} Z^{ij} - z^A \bar z_A }^2 -   \bigl|  \bar \w^{-1} \varepsilon_{ijkl} Z^{ij} Z^{kl}  - \w^{-1}  z^A z_A \bigr|^2 \ee
(see $\cite{nous}$ for the extra-phase factor required in the presence of a non-zero NUT charge). Such charges lye in the $\mathds{R}_+^* \times SL(2,\mathds{R}) \times SO(2,n)$ orbit of $\sl_2$ doublets of linearly independent time-like vectors of $SO(2,n)$,\footnote{where $\Ic(SO(2) \times SO(m)) \cong \scal{ÊSO(2) \times SO(m) } \ltimes \scal{Ê({\bf 2}Ê\otimes {\bf m})^\ord{1} \oplus {\bf 1}^\ord{2}}$, see \cite{nous}.}
\be  \frac{\mathds{R}_+^* \times SL(2,\mathds{R}) \times SO(2,n)}{SO(2) \times SO(n)} \subset \frac{SO(6,2) \times SO(2,n)}{\Ic(SO(2) \times SO(4)) \times SO(n)} \ee
In the limit for which one of the vectors of the doublet becomes null, 
\be \scal{Ê|\w|^2 + |z|^2 - z^A \bar z_A }^2 =  \bigl| 2 \w \bar z - z^A z_A \bigr|^2  \ee
the corresponding black hole has a vanishing horizon area, and the corresponding orbit is 
\be \frac{SL(2,\mathds{R}) \times SO(2,n)}{\mathds{R} \times ISO(n-1)} \subset \frac{SO(6,2) \times SO(2,n)}{\scal{ÊSO(1,1) \times SO(4)} \ltimes  \scal{Ê({\bf 1} \oplus {\bf 4} )^\ord{1} \oplus {\bf 4}^\ord{2} \oplus {\bf 1}^\ord{3}}\times ISO(n-1) } \ee
When the two independent vectors become null,
\beÊ|\w|^2 + |z|^2  = z^A \bar z_A \hspace{20mm}  2 \w \bar z =  z^A z_A  \ee
 the corresponding black hole has one charge associated to the vector multiplets which is saturated, \ie $Êz^A \bar z_A + \sqrt{ (  Êz^A \bar z_A )^2 - |z^A z_A|^2 } = 2 |\w|^2$,
\begin{multline}  \frac{SL(2,\mathds{R}) \times SO(2,n)}{\mathds{R} \times \Ic\scal{ÊSO(1,1) \times SO(n-2)}} \\*\subset \frac{SO(6,2) \times SO(2,n)}{\scal{ÊGL(2,\mathds{R}) \times SO(4) \times SO(n-2)} \ltimes  \scal{Ê{\bf 1}^\ord{-2}Ê \oplus ({\bf 2} \otimes {\bf 4} )^\ord{-1} \oplus ({\bf 2} \otimes ({\bf n-2}))^\ord{1} \oplus {\bf 1}^\ord{2}} } \end{multline}
When the two vectors of the doublet coincide, but remain time-like (or equivalently when one of the vector vanishes), the corresponding black hole is a generic \ft12 BPS black hole ($|z|^2 = |\w|^2$),
\be  \frac{SL(2,\mathds{R}) \times SO(2,n)}{\mathds{R} \times SO(1,n)} \subset \frac{SO(6,2) \times SO(2,n)}{ISO(5,1) \times SO(1,n) } \ee
and when the vector is moreover null, all the charges of the black holes are saturated,
\be z^A \bar z_A = 2 |z|^2 = 2 |\w|^2  \hspace{20mm}  2 \w \bar z =  z^A z_A  \ee
 and the latter would be \ft12 BPS within $\N=8$ supergravity,
\be \frac{SL(2,\mathds{R}) \times SO(2,n)}{IGL(1,\mathds{R}) \times  ISO(1,n-1)} \subset \frac{SO(6,2) \times SO(2,n)}{\mathds{R}_+^* \times ISO(5,1) \times ISO(1,n-1) } \ee
As it is well established, all these solutions can be understood within the $\N=2$ truncation of the $\N=4$ theories corresponding to $\N=2$ supergravity coupled to $1+n$ vector multiplets, with the special K\"{a}hler homogeneous geometry of the coset $SL(2,\mathds{R})/ SO(2) \times SO(2,n)/ (SO(2) \times SO(n))$, which leads after time-like dimensional reduction to the non-linear sigma model over the coset space $SO(4,2+n) / ( SO(2,2) \times SO(2,n))$. The generator ${\bf H}_\frac{1}{4}$ decomposes $\so(2,2)$ as
\be \so(2,2) \cong {\bf 1}^\ord{-4} \oplus \scal{Ê\gl_1 \oplus \sl_2}^\ord{0} \oplus {\bf 1}^\ord{4} \ee
and the coset component of $\so(4,2+n)$ as follows 
\be \so(4,2+n) \ominus \scal{Ê\so(2,2) \oplus \so(2,n)}  \cong \scal{ {\bf 2} \otimes ({\bf 2+n})}^\ord{-2} \oplus \scal{ {\bf 2} \otimes ({\bf 2+n})}^\ord{2} \ee
Note nonetheless that the asymptotic values of the scalar fields are restricted to lie inside the subspace $SL(2,\mathds{R})/ SO(2) \times SO(2,n)/ (SO(2) \times SO(n))$ within the truncated theory.

\subsection{\ft12 BPS solutions}
Using again the `Dirac equation' (\ref{Dirac}), one finds that solutions that preserve eight supersymmetry charges associated to the Killing spinors 
\be \epsilon_\alpha^i + \varepsilon_{\alpha\beta} \Omega^{ij} \epsilon_\beta^j  = 0 \ee
(where $\Omega_{ij}$ is a non-degererate antisymmetric real tensor satisfying $\Omega_{ik} \Omega^{jk} = \delta_i^j$) must have a charge matrix that verifies
\be |\C\rangle = \Scal{Ê\, \w  \, e^{\frac{1}{2}  \Omega_{ij} a^i a^j  } |0\rangle \, ,\, \varrho^A\,   e^{\frac{1}{2}  \Omega_{ij} a^i a^j  }   |0\rangle \, }\ee
with real $\varrho^A$ satisfying $\varrho^A  \varrho_A  \le 2 |\w|^2$. The associated generator ${\bf H}_\frac{1}{2}$ of $\so^*(8)$
\be {\bf H}_\frac{1}{2} \equiv\frac{1}{2}\Scal{Ê \Omega_{ij} a^i a^j - \Omega^{ij} a_i a_j} \ee
defines $|\C\rangle$ as well from the condition ${\bf H}_\frac{1}{2} \, |\C\rangle = 2 |\C\rangle$ and decomposes $\so^*(8)$ as follows,
\be \so^*(8) \cong {\bf 6}^\ord{-2} \oplus \scal{Ê\gl_1 \oplus \su^*(4)}^\ord{0} \oplus {\bf 6}^\ord{2} \ee
such that 
\be \so(8,2+n) \ominus \scal{Ê\so^*(8) \oplus \so(2,n)}  \cong ({\bf 2+n})^\ord{-2} \oplus \scal{ {\bf 6} \otimes ({\bf 2+n})}^\ord{0} \oplus  ({\bf 2+n})^\ord{2} \ee
The \ft12 BPS multi-black solutions thus depend on $2+n$ harmonic functions associated to non-space-like vectors of $SO(2,n)$. Each black hole can be either a generic \ft12 BPS black hole corresponding to a time-like vector ($\varrho^A  \varrho_A  < 2 |\w|^2$) lying in the orbit,
\be \frac{ \mathds{R}_+^* \times SO(2,n)}{SO(1,n)} \subset \frac{SO(6,2) \times SO(2,n)}{ISO(5,1) \times SO(1,n) } \ee
or a black hole with all charges saturated corresponding to a null vector ($\varrho^A  \varrho_A  = 2 |\w|^2$) lying in the orbit,
\be \frac{ SO(2,n)}{ISO(1,n-1)} \subset \frac{SO(6,2) \times SO(2,n)}{\mathds{R}_+^* \times ISO(5,1) \times ISO(1,n-1)} \ee

\section{Non-BPS solutions} 
\label{NBN4}
There are two $SO(6,2)\times SO(2,n)$ orbits of non-BPS spherically symmetric extremal black holes of non-vanishing horizon area. One corresponds to black holes for which the matter charge $\scriptstyle z^A \bar z_A + \sqrt{ (  Êz^A \bar z_A )^2 - |z^A z_A|^2 }$ is saturated, \ie
\be |\w|^4 - |\w|^2 z^A \bar z_A + | z^A z_A |^2 = 0 \ee 
Such black hole are similar to the \ft14 black holes and would be in the same $Spin^*(16)$ orbit of \ft18 BPS black holes within maximal supergravity. The corresponding  multi-black hole solutions are associated to the decomposition of $\so(2,n)$ :
\be \so(2,n) \cong {\bf 1}^\ord{-4} \oplus \scal{  {\bf 2} \otimes ({\bf n-2})}^\ord{-2} \oplus  \scal{\gl_1 \oplusÊ\sl_2 \oplus \so(n-2) }^\ord{0} \oplus \scal{  {\bf 2} \otimes( {\bf n-2})}^\ord{2} \oplus {\bf 1}^\ord{4} \ee
which gives rise to the following decomposition of the coset component of $\so(8,2+n)$,
\be \so(8,2+n) \ominus \scal{Ê\so^*(8) \oplus \so(2,n)}  \cong( {\bf 2} \otimes {\bf 8})^\ord{-2} \oplus \scal{ ({\bf n-2}) \otimes {\bf 8}}^\ord{0} \oplus ( {\bf 2} \otimes {\bf 8})^\ord{2} \ee
Such solutions thus depend on $16$ independent harmonic functions associated to $\sl_2$ doublets of non-space-like vectors of $SO(2,6)$. As in the case of the \ft14 BPS black holes, the black holes associated to a doublet for which one of the $SO(2,6)$ vectors is null have a vanishing horizon area. They are \ft14 BPS if the two vectors are null, and \ft12 BPS if the two vectors moreover coincide. The black holes corresponding to coincident time-like vectors have all their matter charges saturated, \ie $ z^A \bar z_A \pm \sqrt{ (  Êz^A \bar z_A )^2 - |z^A z_A|^2 } = 2 |\w|^2 $, while they do not preserve any supersymmetry. Although they do not preserve any supersymmetry, these solutions correspond to \ft12-BPS solutions of the $\N=2$ supergravity coupled to seven vector multiplets whose bosonic sector defines the consistent truncation of the $\N=4$ theory obtained by disregarding all the vector multiplets whose associated charges vanish on the horizons. The corresponding non-linear sigma model of the latter $\N=2$ truncation is defined over
\be SO(8,4) / ( SO(6,2) \times SO(2,2) )  \subset SO(8,2+n) / ( SO(6,2) \times SO(2,n)) \ee

The other $SO(6,2)\times SO(2,n)$ orbit of non-BPS spherically symmetric extremal black holes correspond to black holes for which none of the charges are saturated, and that would not be BPS within maximal supergravity. They are associated to the following decomposition of $\so(6,2) \oplus \so(2,n)$,
 \begin{multline} \so(6,2) \oplus \so(2,n) \cong \scal{Ê{\bf 6}_- \oplus {\bf n}_+}^\ord{-2} \\* \oplus  \scal{\gl_1 \oplusÊ\so(1,1) \oplus \so(5,1) \oplus \so(1,n-1) }^\ord{0} \oplus \scal{Ê{\bf 6}_+ \oplus {\bf n}_-}^\ord{2}  \label{NBgrad}\end{multline}
 which gives rise to the decomposition of the coset component of $\so(8,2+n)$,
\begin{multline} \so(8,2+n) \ominus \scal{Ê\so^*(8) \oplus \so(2,n)}  \\* \cong {\bf 1}^\ord{-4} \oplus  \scal{Ê{\bf n}_-\oplus {\bf 6}_+}^\ord{-2} \oplus \scal{ {\bf 1}_{\scriptscriptstyle --} \oplus {\bf 6} \otimes {\bf n}  \oplus {\bf 1}_{\scriptscriptstyle ++}}^\ord{0} \oplus \scal{Ê{\bf n}_+ \oplus {\bf 6}_-}^\ord{2}  \oplus {\bf 1}^\ord{4} \label{NBcoset} \end{multline}
where the indices $\pm$ indicate the weight with respect with $\so(1,1)$. The component $\scal{Ê{\bf n}_+ \oplus {\bf 6}_-}^\ord{2}  \oplus {\bf 1}^\ord{4}$ defines an abelian sub-algebra $\mathds{R}^{n+6+1}$, and one has associated multi-black holes solutions depending on $n+7$ harmonic functions. 

The grading (\ref{NBgrad}) associated to the non-BPS extremal solutions can be defined for example by the $\so(6,2) \oplus \so(2,n)$ generator (satisfying ${\bf H}^3 = 4 {\bf H}$),
\be {\bf H} \equiv  \left( \begin{array}{cccc} \,\hspace{2mm}  0 \hspace{2mm} \, & \,\hspace{2mm}    0 \hspace{2mm} \, & \,  
\, \baa \hat{Q}    \, & \, -  \ba \hat{p}  \, \\
\,  0 \, & 0 & - \ba \hat{p}  &  \, \baa \hat{Q}  \\  \, \baa \hat{Q}  &   
\ba \hat{p}  & 0 &  0 \\
 \ba \hat{p} &  \, \baa \hat{Q}   &   0 & 0 \end{array} \right) 
\ee 
where the hats mean that the vectors are normalised, such that $ \baa \hat{Q} ^2 = 1$ and $ \ba \hat{p} ^2 = - 1$ according to the $Spin(6,n)$ Clifford algebra. Of course the whole $SO(6,2) \times SO(2,n)$ orbit of this generator defines equivalent graded decompositions. According to the decomposition (\ref{NBcoset}), there is one single charge matrix of grade four with respect with this generator, which corresponds to the pure NUT maximally supersymmetric charge matrix 
 \be 
\C^\ord{4} = N  \left( \begin{array}{cccc} \, \hspace{2mm}   \baa \hat{Q} \ba \hat{p} \hspace{2mm}   \, & \,   \hspace{2mm} \hat{Q} \hspace{2mm} \, & \,\hspace{2mm}  \ba \hat{p} \hspace{2mm} \, & \,  \hspace{2mm} -   \baa \hat{Q} \hspace{2mm}  \, \\
\,  \hat{Q} \, &  \baa \hat{Q} \ba \hat{p}  &  -  \baa \hat{Q} &  \, \ba \hat{p} \\  \,\ba \hat{p} &   
 \baa \hat{Q}   &  \baa \hat{Q} \ba \hat{p}&  \hat{Q} \\
   \baa \hat{Q}  &  \ba \hat{p} &   \hat{Q} &  \baa \hat{Q} \ba \hat{p}  \end{array} \right) 
\ee 
The charges matrix of grade two depend on an $SO(1,5)$ and an $SO(1,n-1)$ vector which decompose as $\ba p$ and a vector $\baaaP$ of $SO(6)$ orthogonal to $\baa Q$, and as $\baa Q$ and a vector $\ba q$ of $SO(n)$ orthogonal to $\ba p$, respectively. It is given by 
\be \C^\ord{2} = \left( \begin{array}{cccc} \,  Q + p + \baa \hat{Q} \ba q + \baaaP \ba \hat{p}Ê\, & \,   0 \, & \, -  
\, \baa Q  + \ba q \, & \, \ba p -   \baaaP \, \\
\,  0 \, & -Q - p + \baa \hat{Q} \ba q + \baaaP \ba \hat{p} & - \ba p -  \baaaP &  \, \baa Q + \ba q \\  \, \baa Q  + \ba q &   
- \ba p + \baaaP   & -Q + p + \baa \hat{Q} \ba q + \baaaP \ba \hat{p} &   0 \\
   \baaaP  + \ba p & -  \, \baa Q  + \ba q &  0 & Q - p + \baa \hat{Q} \ba q + \baaaP \ba \hat{p} \end{array} \right)  \label{NBcharge}
\ee 
which is linear in the norms $Q$ and $p$ of $\baa Q$ and $\ba p$, respectively, and in $\ba q$ and $\baaaP$. We see that having chosen ${\bf H}$ such that the grade two charge matrix have a vanishing NUT charge, the grade four matrix has a vanishing mass and a non-zero NUT charge. If one wants to consider asymptotically Minkowski multi-black holes solutions one has therefore to restrict to charge matrices lying in the grade two component. Note nevertheless that the black holes of charge matrix of the form $\C^\ord{2}$ would still satisfy a no force property inside the maximally supersymmetric Taub--NUT space-times associated to black holes of charge matrix $\C^\ord{4}$.

Black holes carrying a charge matrix $\C^\ord{2}$ have a mass 
\be M = Q + p \ee
It is saturated, if and only if either $P = p$, in which case the solution is \ft14 BPS, or $q=Q$, in which case the solution is still non-BPS within $\N=4$ supergravity. The conditions for such solutions to be regular are thus,
\be q \le Q \hspace{20mm} P \le p \ee
which mean that the corresponding vector of $SO(1,n-1)$ and $SO(1,5)$ have to be non-space-like. As we will see in the next section, the horizon area of such black hole is given by 
\be A_\mathscr{H} = 16 \pi \sqrt{ ( Q^2 - q^2) ( p^2 - P^2 )} \ee
which is nothing else than the square root of $-\lozenge( Z) > 0$. 

The generic black holes thus correspond to combinations of time-like vectors of $SO(1,5)$ and $SO(1,n-1)$ (\ie such that  $q < Q$ and $ P < p$), which lie in the orbit
\be \frac{ \mathds{R}_+^* \times SO(1,1) \times SO(1,5) \times SO(1,n-1)}{ SO(5) \times SO(n-1)}Ê\subset \frac{ SO(6,2) \times SO(2,n)}{ \mathds{R} \times ISO(5) \times ISO(n-1)} \ee
If the vector of $SO(1,5)$ is null (\ie $P=p$), the corresponding black hole is \ft14 BPS (because then $M = Q+P$) and has a vanishing horizon area.
\begin{multline} \frac{  SO(1,1) \times SO(1,5) \times SO(1,n-1)}{ ISO(4) \times SO(n-1)}\\*Ê\subset \frac{ SO(6,2) \times SO(2,n)}{\scal{ÊSO(1,1) \times SO(4) \times SO(n-1)}\ltimes \scal{({\bf 1}\oplus {\bf 4} \oplus {\bf n-1})^\ord{1} \oplus {\bf 4}^\ord{2} \oplus {\bf 1}^\ord{3}}} \end{multline}
If the $SO(1,n-1)$ vector is null (\ie $Q= q$), the corresponding black hole has a saturated matter charge  ($M = q + p$) and a vanishing horizon area. 
\begin{multline} \frac{  SO(1,1) \times SO(1,5) \times SO(1,n-1)}{ SO(5) \times ISO(n-2)}\\*Ê\subset \frac{ SO(6,2) \times SO(2,n)}{\scal{ÊSO(1,1) \times SO(5) \times SO(n-2)}\ltimes \scal{({\bf 1}\oplus {\bf 5} \oplus {\bf n-2})^\ord{1} \oplus {\bf n-2}^\ord{2} \oplus {\bf 1}^\ord{3}}} \end{multline}
When both vectors are null the corresponding black hole is \ft14 BPS and has a saturated matter charge (\ie $M = Q + P= q+p$). 
\begin{multline} \frac{  SO(1,1) \times SO(1,5) \times SO(1,n-1)}{ ISO(4) \times ISO(n-2)}\\*Ê\subset \frac{ SO(6,2) \times SO(2,n)}{\scal{ÊGL(2,\mathds{R}) \times SO(4) \times SO(n-2)}\ltimes \scal{{\bf 1}^\ord{-2} \oplus ({\bf 2}\otimes {\bf 4})^\ord{-1}  \oplus ({\bf 2} \otimes ({\bf n-2}))^\ord{1} \oplus {\bf 1}^\ord{2} }} \end{multline}
If the $SO(1,5)$ vector vanishes ($P=p=0$) the corresponding black hole is \ft12 BPS ($M=Q$),\footnote{Recall that $P= 0$ implies that $Z_{ij}$ is complex self-dual, and thus that $|\z_\un| = |\z_\deux| = Q$.} 
\be \frac{  SO(1,1) \times SO(1,5) \times SO(1,n-1)}{ SO(1,5) \times SO(n-1)}Ê\subset \frac{ SO(6,2) \times SO(2,n)}{ ISO(5,1) \times SO(1,n)} \ee
and if the $SO(1,n-1)$ vector vanishes ($Q=q=0$), the corresponding black hole has its two matter charges saturated ($M=p$). 
\be \frac{  SO(1,1) \times SO(1,5) \times SO(1,n-1)}{ SO(5) \times SO(1,n-1)}Ê\subset \frac{ SO(6,2) \times SO(2,n)}{ SO(6,1) \times ISO(1,n-1)} \ee
The black holes associated to a null vector of either $SO(1,5)$ or $SO(1,n-1)$ and a vanishing vector of the other group both correspond to \ft12 BPS black holes with all the charges saturated, that would be \ft12 BPS within maximal supergravity (then either $M=Q=q$ and $P=p=0$ or $M=P=p$ and $Q=q=0$).
\begin{multline} \frac{  SO(1,1) \times SO(1,5) \times SO(1,n-1)}{ ISO(4) \times SO(1,n-1)}Ê\cup \frac{  SO(1,1) \times SO(1,5) \times SO(1,n-1)}{ SO(1,5) \times ISO(n-2)}  \cup \mathds{R}\\* \subset \frac{ SO(6,2) \times SO(2,n)}{ \scal{ÊSO(1,1) \times SO(5,1) \times SO(1,n-1)} \ltimes \scal{Ê{\bf 6}^\ord{1} \oplus {\bf n}^\ord{-1}}} \end{multline}
One can check the nilpotency conditions stated in \cite{nous} associated to the amount of saturated charges in each case. 

\section{A class of explicit solutions}
\label{solution}
Let us illustrate the abstract formalism we introduced in this paper by using it to derive the explicit non-BPS multi-black hole solutions of $\N=4$ supergravity coupled to $n\ge 2$ vector multiplets. As a matter of fact, the formulas of this section generalise trivially to gravity coupled to scalar fields lying in the homogeneous space $SL(2,\mathds{R})/ SO(2) \cup SO(n,m) / ( SO(n) \times SO(m))$ and abelian vector fields in the vector representation of $SO(n,m)$, but we will restrict ourselves to $\N=4$ supergravity for the sake of clarity. For simplicity we will restrict ourselves to solutions with trivial moduli in the asymptotic region (\ie with $\V_\zero = \mathds{1}$). The general solutions can be obtained straightforwardly by acting on the solutions with the four-dimensional duality group $SL(2,\mathds{R}) \times SO(6,n)$.

The general solutions associated to charges of the form (\ref{NBcharge})  in the symmetric gauge $ \V = \exp(- \sum_\m \Ha_\m \C^\ord{2}_\m)$ is easy to get, but it does not exhibit the expressions of the four-dimensional fields. For this purpose one needs to write down the coset representative $\V$ in the parabolic gauge (\ref{KKa}). One can carry out this rotation by multiplying $\V$ to the right by an element of $SO(6,2) \times SO(2,n)$ of the form
\be u =  \left(\begin{array}{cccc} \hspace{2mm} a_+ b_+ \hspace{2mm}& \hspace{2mm} \alpha_- \beta_- \hspace{2mm}& \hspace{2mm}  \alpha_-  b_+  \hspace{2mm}& \hspace{2mm} - a_+ \beta_-  \hspace{2mm}  \\
 \hspace{2mm} \alpha_+ \beta_+  \hspace{2mm} & \hspace{2mm}  a_- b_-  \hspace{2mm} & \hspace{2mm} - a_- \beta_+ \hspace{2mm} & \hspace{2mm} \alpha_+ b_-  \hspace{2mm} \\    \alpha_+b_+   &  a_- \beta_-   &  a_- b_+  &  - \alpha_+ \beta_-   \\  a_+ \beta_+   &  \alpha_- b_- & - \alpha_- \beta_+   & a_+ b_- \end{array} \right) \ee
 where the Clifford algebra elements $a_\pm ,\, b_\pm,\, \alpha_\pm $ and $ \beta_\pm$ are defined in function of the two orthogonal vectors of $SO(6)$, $\baa X$ and $\baa Y$, and the two orthogonal vectors of $SO(n)$,  $\ba{\rm x}$ and $\ba {\rm y}$, as follows
 \be \begin{split}
 a_\pm &\equiv \frac{ ( 1+{\rm x}) ( 1 + X ) \pm \baa X \ba {\rm x}}{\sqrt{ ( 1 + 2 {\rm x} ) ( 1 + 2 X )}} \\
  b_\pm &\equiv \frac{ ( 1+{\rm y}) ( 1 + Y ) \pm \baa Y \ba {\rm y}}{\sqrt{ ( 1 + 2 {\rm y} ) ( 1 + 2 Y )}}
 \end{split}\hspace{20mm}\begin{split}
  \alpha_\pm &\equiv  \frac{ ( 1+{\rm x}) \baa X \pm ( 1 + X)  \ba {\rm x}}{\sqrt{ ( 1 + 2 {\rm x} ) ( 1 + 2 X )}} \\
 \beta_\pm &\equiv  \frac{  ( 1 + Y)  \ba {\rm y} \pm ( 1+{\rm y}) \baa Y  }{\sqrt{ ( 1 + 2 {\rm y} ) ( 1 + 2 Y )}}
 \end{split}\ee
 with $X,\, Y,\, {\rm x}$ and ${\rm y}$ being the norm of these vectors, and $\baa X,\, \baa Y,\, \ba {\rm x}$ and $\ba {\rm y}$ being parallel to $\baa Q,\, \baaaP,\, \ba q$ and $\ba p$, respectively. To simplify notations we will refer to $\exp(-\C^\ord{2})$ rather than $\V$, the latter being obtained trivially from the former by substituting the harmonic functions to the corresponding charges. ${\C^\ord{2}}^3= 0$ and thus $\exp(-\C^\ord{2})$ takes the simple form 
 \be \exp(-\C^\ord{2}) =  \left(\begin{array}{cccc} \hspace{2mm} \mathcal{Q}_- \mathcal{P}_- \hspace{2mm}& \hspace{2mm} \chi_- \pi_- \hspace{2mm}& \hspace{2mm} \chi_-   \mathcal{P}_- \hspace{2mm}& \hspace{2mm} - \mathcal{Q}_- \pi_-  \hspace{2mm}  \\
 \hspace{2mm} \chi_+ \pi_+  \hspace{2mm} &\hspace{2mm}  \mathcal{Q}_+ \mathcal{P}_+  \hspace{2mm} & \hspace{2mm}  \mathcal{Q}_+ \pi_+ \hspace{2mm} & \hspace{2mm}  - \chi_+ \mathcal{P}_+ \hspace{2mm} \\   -  \chi_+  \mathcal{P}_- &  \mathcal{Q}_+ \pi_-   &  \mathcal{Q}_+ \mathcal{P}_-  &  \chi_+ \pi_-   \\  -\mathcal{Q}_- \pi_+   & \mathcal{P}_+ \chi_-  &  \chi_- \pi_+   & \mathcal{Q}_- \mathcal{P}_+\end{array} \right) \ee
where
\be\begin{split}
 \mathcal{Q}_\pm &\equiv 1 \pm Q - \baa \hat Q \ba q  \\
 \chi_\pm &\equiv  \baa Q \pm \ba q
 \end{split}\hspace{10mm}\begin{split}
  \mathcal{P}_\pm &\equiv  1 \pm p - \baaaP \ba \hat{p}  \\
 \pi_\pm &\equiv  \ba p \pm \baaaP
 \end{split}\ee
The conditions for $\exp(- \C^\ord{2}) \cdot u $ to be of the form (\ref{KKa}),\footnote{Note that $\mathcal{Q}_\pm$ commute with both $b_\pm$ and $\beta_\pm$ and that $\chi_\pm$ commute with $b_\pm$ and anticommute with $\beta_\pm$; and so do respectively $\mathcal{P}_\pm$ and $\pi_\pm$ with respect with $a_\pm$ and $\alpha_\pm$.} are $\mathcal{Q}_+ \alpha_+ = \chi_+ a_+$ and $\mathcal{P}_+ \beta_+ = \pi_+ b_+$, which read
\be\begin{split}
 \scal{ 1 + Q - q + X}\,  \ba {\rm x}  &= \ba q  \\
(Ê1 + {\rm x}  )\,  \scal{ \baa X - \baa Q}   &= -  q {\rm x}\,  \baa \hat{Q} 
 \end{split}\hspace{10mm}\begin{split}
 \scal{ 1 + p - P + {\rm y} }\,  \baa Y  &= \baa P  \\
(Ê1 + Y  ) \, \scal{ \ba {\rm y} - \ba p}   &=  - P Y \, \ba \hat{p}  \end{split}\ee
and have as relevant solutions 
\be\begin{split}
 {\rm x}   &= \frac{ \sqrt{ 1 + 2 Q + 2 q} - \sqrt{1 + 2Q - 2q}}{2 \sqrt{ 1 + 2 Q - 2 q}}  \\
Y   &= \frac{ \sqrt{ 1 + 2 p + 2 P} - \sqrt{1 + 2p - 2P}}{2 \sqrt{ 1 + 2 p - 2 P}} 
  \end{split}\hspace{10mm}\begin{split}
X &=  Q - \frac{  \sqrt{ 1 + 2 Q + 2 q} - \sqrt{1 + 2Q - 2q}}{ \sqrt{ 1 + 2 Q + 2 q} + \sqrt{1 + 2Q - 2q}} q \\
{\rm y}  &=  p - \frac{  \sqrt{ 1 + 2 p + 2 P} - \sqrt{1 + 2p - 2P}}{ \sqrt{ 1 + 2 p + 2 P} + \sqrt{1 + 2p - 2P}} P
  \end{split}\ee

In order to write down the general solution, we define a basis of $5$ normed $SO(6)$ vectors $\baaahP^\a$ orthogonal to $\, \baa \hat{Q}$, as well as a basis of  $n-1$ normed $SO(n)$ vectors  $\ba \hat{q}^\m$ orthogonal to $\ba \hat{p}$, and the following $6+n$ harmonic functions 
\be\begin{split}
 \Ha_\zero &\equiv 1 + 2 \sum_{\scriptscriptstyle A}  \frac{Q_{\scriptscriptstyle A}}{ | x - x_{\scriptscriptstyle A}|} \\
 \Ka_\zero &\equiv 1 + 2 \sum_{\scriptscriptstyle A}  \frac{p_{\scriptscriptstyle A}}{ | x - x_{\scriptscriptstyle A}|} 
 \end{split}\hspace{20mm}\begin{split}
  \Ha_\m &\equiv  2 \sum_{\scriptscriptstyle A}  \frac{q^\m_{\scriptscriptstyle A}}{ | x - x_{\scriptscriptstyle A}|} \\
 \Ka_\a &\equiv 2 \sum_{\scriptscriptstyle A}  \frac{P^\a_{\scriptscriptstyle A}}{ | x - x_{\scriptscriptstyle A}|} 
 \end{split}\ee
 which verify for each pole $x_{\scriptscriptstyle A}$ that 
 \be { Q_{\scriptscriptstyle A}}^2  \ge \sum_{\m= 1}^{n-1} { q^\m_{\scriptscriptstyle A} }^{\, 2} \hspace{20mm}  { p_{\scriptscriptstyle A}}^2  \ge \sum_{\a=1}^{5} { P^\a_{\scriptscriptstyle A} }^{\, 2} \ee
The fields of the Kaluza--Klein ansatz (\ref{KKa}) are given by 
\begin{gather} H = \Scal{ {\Ha_\zero}^2 - \sum  { \Ha_\m }^2}^{-\frac{1}{2}} \Scal{ {\Ka_\zero}^2 - \sum  { \Ka_\a }^2}^{-\frac{1}{2}}  \CR
\bar H = \Scal{ {\Ha_\zero}^2 - \sum  { \Ha_\m }^2}^{\frac{1}{2}} \Scal{ {\Ka_\zero}^2 - \sum  { \Ka_\a }^2}^{-\frac{1}{2}} \CR
U^\zero \equiv \frac{1}{2} \{Ê\baa \hat{Q} , \baa U \} = 1 - \frac{ \Ha_\zero}{ {\Ha_\zero}^2 - \sum  { \Ha_\m }^2}  \hspace{10mm}
U^\m \equiv    \frac{1}{2} \{Ê\ba \hat{q}^\m , \baa U \} =   \frac{ \Ha_\m}{ {\Ha_\zero}^2 - \sum  { \Ha_\n }^2}  \CR
A^\zero \equiv  \frac{1}{2} \{Ê\ba \hat{p} , \baaa A \} = 1 - \frac{ \Ka_\zero}{ {\Ka_\zero}^2 - \sum  { \Ka_\a }^2}  \hspace{10mm}
A^\a \equiv \frac{1}{2} \{Ê \baaahP^\a , \baaa A \} =    \frac{ \Ka_\a}{ {\Ka_\zero}^2 - \sum  { \Ka_\b }^2}  \CR
v = \frac{  \left(Ê\Ha_+ -  \frac{ 2 \sum_{}  \Ha_\m \bahQ \bahq^\m }{ \Ha_+} \right)   \left(Ê\Ka_+ -  \frac{ 2 \sum_{}    \Ka_\a \baahP^\a  \, \bahp }{ \Ka_+} \right) }{ 4  \Scal{ {\Ha_\zero}^2 - \sum  { \Ha_\n }^2}^{\frac{1}{4}} \Scal{ {\Ka_\zero}^2 - \sum  { \Ka_\b }^2}^{\frac{1}{4}} }
\end{gather}
 where $v$ is in the symmetric gauge (\ie $\ln v \in \so(6,n) \ominus ( \so(6) \oplus \so(n))$) and 
 \bea \Ha_+ &\equiv& \left(  \Ha_\zero + \Scal{ Ê\sum  {\Ha_\m}^2}^{\frac{1}{2}}\right)^\frac{1}{2} + \left(  \Ha_\zero - \Scal{ Ê\sum  {\Ha_\m}^2}^{\frac{1}{2}}\right)^\frac{1}{2} \CR
 \Ka_+ &\equiv& \left(  \Ka_\zero + \Scal{ Ê\sum  {\Ka_\a}^2}^{\frac{1}{2}}\right)^\frac{1}{2} + \left(  \Ka_\zero - \Scal{ Ê\sum  {\Ka_\a}^2}^{\frac{1}{2}}\right)^\frac{1}{2} \eea
 and all the other fields are trivially zero. 

One computes easily that in the vicinity of a pole $x_{\scriptscriptstyle A}$ of the harmonic functions, the function $H$ defining the metric behaves as
\be H = \frac{Ê | x - x_{\scriptscriptstyle A}|^2 }{ 4 \sqrt{Ê\scal{ {Q_{\scriptscriptstyle A}}^2  -  \sum { q^\m_{\scriptscriptstyle A} }^{\, 2} } \scal{Ê  { p_{\scriptscriptstyle A}}^2  - \sum{ P^\a_{\scriptscriptstyle A} }^{\, 2}    } }}  + \mathcal{O}\scal{Ê | x - x_{\scriptscriptstyle A}|^4} \ee
such that the corresponding horizon area is
\be A_{\mathscr{H}_{A}} = 16 \pi \sqrt{Ê\left(  {Q_{\scriptscriptstyle A}}^2  -  \sum { q^\m_{\scriptscriptstyle A} }^{\, 2}\right)\left(Ê  { p_{\scriptscriptstyle A}}^2  - \sum{ P^\a_{\scriptscriptstyle A} }^{\, 2}    \right) } \ee
as stated in the preceding section. If $ {Q_{\scriptscriptstyle A}}^2$ was strictly inferior to $ \sum { q^\m_{\scriptscriptstyle A} }^{\, 2}$, the function $H$ would diverge at a positive value of $| x - x_{\scriptscriptstyle A}|$, and the solution would exhibit a naked singularity. For example, $H$ would diverge at $r = 2 ( q-Q)$ in the case of a spherically black hole. The discussion is equivalent for  ${ p_{\scriptscriptstyle A}}^2  < { P^\a_{\scriptscriptstyle A} }^{\, 2}$.
 
The most general solution of this kind can straightforwardly be obtained by acting with $SO(6,2) \times SO(2,n)$ on the coset representative (\ref{KKa}). For example, one can generate solutions with a non-trivial axion field by an $SO(2)$ rotation
\bea
\bar H (\alpha) &=&  \frac{Ê\Scal{ {\Ha_\zero}^2 - \sum  { \Ha_\m }^2}^{\frac{1}{2}} \Scal{ {\Ka_\zero}^2 - \sum  { \Ka_\a }^2}^{\frac{1}{2}} }{ \cos^2 \alpha Ê \Scal{ {\Ka_\zero}^2 - \sum  { \Ka_\a }^2} + \sin^2 \alpha \Scal{ {\Ha_\zero}^2 - \sum  { \Ha_\n }^2} } \CR
\bar B(\alpha) &=& \frac{1}{2} \sin 2 \alpha \, \frac{ {\Ka_\zero}^2 - \sum  { \Ka_\a }^2 - {\Ha_\zero}^2 + \sum  { \Ha_\m }^2 }{ \cos^2 \alpha Ê \Scal{ {\Ka_\zero}^2 - \sum  { \Ka_\a }^2} + \sin^2 \alpha \Scal{ {\Ha_\zero}^2 - \sum  { \Ha_\n }^2} } \CR
\baa U(\alpha) &=& \cos \alpha \, \baa \hat {Q} - \sin \alpha \, \ba \hat{p} - \cos \alpha \, \frac{ \Ha_\zero \,  \baa \hat{Q} + \sum \Ha_\m \,  \ba \hat{q}^\m }{ {\Ha_\zero}^2 - \sum  { \Ha_\n }^2} + \sin \alpha \,  \frac{ \Ka_\zero \, \ba \hat{p} + \sum \Ka_\a \,  \baaahP^\a }{ {\Ka_\zero}^2 - \sum  { \Ka_\b }^2} \CR
\baaa A(\alpha) &=& - \cos \alpha \, \ba \hat{p} - \sin \alpha \, \baa \hat {Q} +  \cos \alpha \,  \frac{ \Ka_\zero \, \ba \hat{p} + \sum \Ka_\a \,  \baaahP^\a }{ {\Ka_\zero}^2 - \sum  { \Ka_\b }^2}  + \sin \alpha \, \frac{ \Ha_\zero \,  \baa \hat{Q} + \sum \Ha_\m \,  \ba \hat{q}^\m }{ {\Ha_\zero}^2 - \sum  { \Ha_\n }^2} 
\eea
There is still one missing free parameter for the most general asymptotically Minkowski solution of this type with trivial moduli (\ie with $\V_0 = \mathds{1}$), which can be generated by the nilpotent generator of grade $-2$ of $\mathfrak{so}(6,2) \oplus \so(2,n)$
\be \left( \begin{array}{cccc} \, \hspace{2mm}  0 \hspace{2mm}   \, & \,   \hspace{2mm} 0  \hspace{2mm} \, & \,\hspace{2mm}  \ba \hat{p} \hspace{2mm} \, & \,  \hspace{2mm}    \baa \hat{Q} \hspace{2mm}  \, \\
\,  0 \, & 0 &  \hspace{-1mm} -  \baa \hat{Q} &  \, \hspace{-1mm} - \ba \hat{p} \\  \, \hspace{-1mm} - \ba \hat{p} &   
\hspace{-1mm} -  \baa \hat{Q}   &  0 &  \hspace{-1mm} -2 \\
   \baa \hat{Q}  &  \ba \hat{p} &   2 & 0   \end{array} \right)^3 = 0 
\ee 
Asymptotically Taub--NUT space-times of this kind also exist of course, and can be obtained by acting with the Ehlers $SO(2)$. Then, all the $Q_{\scriptscriptstyle A} + p_{\scriptscriptstyle A}$ have to be integral multiplier of a given fundamental charge in order to avoid Dirac--Misner string singularities \cite{instantons,Komar}.

Note that although these solutions can be embedded into maximal supergravity for $n\le 6$, they do not define the most general non-BPS multi-black holes solutions of this kind within maximal supergravity, which would depend on $28$ independent harmonic functions and not only $12$ (for $n=6$).

Let us consider the case of a spherically symmetric black hole. It is interesting to compute the scalar dependent combinations of the charges generalising the central charges (by including the charges associated to the vector multiplet) on the horizon $\mathscr{H}$. We assume for this purpose that the  horizon has a non-vanishing horizon area. 
\bea \scal{Ê\bar H^{-\frac{1}{2}} \, v^{-1}} \big|_{\mathscr{H}} Ê\scal{Ê\baa Q + \ba q }   \scal{Êv }\big|_{\mathscr{H}}  &=& \sqrt[4]{ \scal{ÊQ^2 - q^2 }Ê\scal{ p^2 - P^2 } } \,\,  \baa \hat{Q} \CR
\scal{Ê\bar H^{\frac{1}{2}} \, v^{-1}} \big|_{\mathscr{H}} Ê\scal{Ê\ba p + \baaaP }   \scal{Êv }\big|_{\mathscr{H}}  &=& \sqrt[4]{ \scal{ÊQ^2 - q^2 }Ê\scal{ p^2 - P^2 } } \,\,  \ba \hat{p}  \eea
These charges are thus uniquely determined by the $\so(6,2)\oplus \so(2,n)$ generator ${\bf H}$ characterising the nilpotent orbit of the Noether charge, and by the horizon area. And inversely, the expression of the `generalised central charges' at the horizon determine uniquely the generator ${\bf H}$ and the horizon area. Note that this is valid for any asymptotic value of the scalar fields since the `generalised central charges'  at the horizon do not depend on them because of the attractor mechanism phenomena \cite{attractors}.
\section{Higher order orbits}
\label{Evidence}
In principle one could have more general multi-black holes solutions associated to higher order orbits. Indeed, as explained in \cite{PapaMajumdar}, any grading associated to a nilpotent orbit which generic representative vanish at the sixth power in the adjoint representation (\ie ${\ad_{\bf E}}^6$) defines a linear space $\mathfrak{n}^\ord{2} \cong \bigoplus_{p\ge 2} ( \g - \h^* )^\ord{p} $ of elements satisfying equations (\ref{tricom}). For example, consider that two charges $\C_\un$ and $\C_\deux$ define regular spherically symmetric black holes, such that the linear combination ${\bf E}(\alpha) = \alpha \C_\un + (1- \alpha ) \C_\deux$ does not satisfy ${\bf E}(\alpha)^3 = 0$, but satisfies nonetheless ${\ad_{{\bf E}(\alpha)}}^6 = 0$ such that equation (\ref{tricom}) is satisfied. Then, one would have regular Papapetrou--Majumdar solutions of a more general type than the one discussed in the preceding sections. Nevertheless, it was argued in \cite{PapaMajumdar} that solutions associated to higher order orbits always carry naked singularities. We are now going to prove this proposition within $\N=4$ supergravity coupled to $n$ vector multiplets. We will then provide some strong evidence that it is also the case in maximal supergravity.

We recall that the regular generic spherically symmetric extremal black holes (\ie with a non-vanishing horizon area) carry a Noether charge which isotropy subgroup of $SO(6,2) \times SO(2,n)$ is a contracted form of $SO(6) \times SO(2) \times SO(n)$ \cite{nous}. This comes from the fact that such black holes appear as particular limit of regular non-extremal spherically black holes which all lie in the $SO(6,2) \times SO(2,n)$ orbit of a Schwarzschild solution \cite{Maison}, and which therefore carry a Noether charge of isotropy subgroup $SO(6) \times SO(2) \times SO(n)$. More generally, the isotropy subgroups of Noether charges associated to regular spherically symmetric black holes have been classified in \cite{nous}. 

As we are going to see, whenever the linear combination ${\bf E}(\alpha) $ of two Noether charges satisfying ${\C_\un}^3 = {\C_\deux}^3 = 0$ lies in the intersection of a higher order orbit with the coset component $\so(8,2+n) \ominus \scal{Ê\so(6,2) \oplus \so(2,n)}$, their isotropy subgroup is always such that they correspond to singular black holes without horizon.

The nilpotent $O(10+n,\mathds{C})$-orbits of $\so(10+n,\mathds{C})$ are in one to one correspondence with the partitions of $10+n$ carrying an even number of each even integer involved in the partition \cite{coadjoint}.  For example, $(4)^{2}(3)^1(2)^2(1)^1$ states for the partition $16 = 4+4+3+2+2+1$. The partitions $(2)^{2s} (1)^{10+n-4s}$ are associated to nilpotent orbits of dimension $2s(9+n-2s)$ for which ${\ad_{E}}^3 = 0$,  and ${E}^{1+s}=0$ in the spinor representation. The partitions $(3)^{1+p} (2)^{2s} (1)^{7+n - 3p - 4s}$ are associated to nilpotent orbits of dimension $p(16+2n-3p-2s) + 2s(8+n-p-2s)$ for which ${\ad_{E}}^5 = 0$,  and ${E}^{2+p+s} = 0$ in the spinor representation. They are all the orbits we are interested in because the partitions involving higher integers all satisfy ${\ad_{E}}^6 \ne 0$.\footnote{To see this, one computes that the elements of the orbits associated to the partitions $(4)^2(1)^{2+n}$ and $(5)(1)^{5+n}$ only vanish at the seventh power in the adjoint representation. The result for the other orbits then follows from the closure ordering of the nilpotent orbits \cite{coadjoint,Rousset}.} The characteristic equation ${E}^3 = 0$ restricts to the orbits of partition $(3)^p(2)^{2s}(1)^{10+n-3p-4s}$ with $p+s \le 2$, which correspond to regular spherically symmetric extremal black holes. The nilpotent orbits associated to generic extremal solutions correspond to the partition $(3)^2 (1)^{4+n}$, and so nilpotent linear combinations of such elements which vanish at the sixth power in the adjoint representation lye in a nilpotent orbit associated to a partition $(3)^{2+p} (2)^{2s} (1)^{4+n - 3p - 4s}$ \cite{SOpqstrat}. For a non-zero $s$, the ninth-graded decomposition associated to such orbit is
\begin{multline} \so(8,2+n) \cong \overline{ \bf \sfrac{1}{2}(p+2)(p+1)}^\ord{-4} \oplus \scal{Ê \overline{({\bf p+2})} \otimes \overline{\bf 2s}}^\ord{-3}_- \\* \oplus \scal{ÊÊ \overline{({\bf p+2})} \otimes({\bf 6 +n - 2p - 4s}) \oplus \overline{\bf s ( 2s-1)}_{--}  }^\ord{-2}\\*  \oplus   \scal{Ê \overline{({\bf p+2})} \otimes {\bf 2s}_+ \oplus \overline{\bf 2s} \otimes ({\bf 6 +n - 2p - 4s})_- }^\ord{-1} \\* \oplus  \scal{\gl_1 \oplusÊ\so(1,1) \oplus \sl_{2+p} \oplus \sl_{2s} \oplus \so(6-p-2s,n-p-2s)}^\ord{0} \\*   \oplus \scal{Ê ({\bf p+2}) \otimes \overline{\bf 2s}_- \oplus {\bf 2s} \otimes ({\bf 6 +n - 2p - 4s})_+ }^\ord{1}\\* \oplus   \scal{ÊÊ{({\bf p+2})} \otimes({\bf 6 +n - 2p - 4s}) \oplus { \bf s ( 2s-1)}_{++}  }^\ord{2} \\* \oplus \scal{Ê{({\bf p+2})} \otimes {\bf 2s}}^\ord{3}_+ \oplus {\bf \sfrac{1}{2}(p+2)(p+1)}^\ord{4} \label{geneHO}\end{multline} 
Where the $\pm$ subscripts state for the corresponding representation of $\so(1,1)$. The corresponding representative $E$ are generic elements of the grade two component, that involve a $(p+2)$-plet of orthogonal non-null vectors of $SO(6-p-2s,n-p-2s)$ and a non-degenerated component in the ${\bf  s ( 2s-1)}$ of $SL(2s,\mathds{R})$, which altogether are left invariant by a subgroup
\be SO(p_+, 2+p-p_+) \times Sp(2s,\mathds{R}) \times SO(6-p_+-p-2s,n-2+p_+-2p-2s) \ee
of the Levy subgroup (grade zero component) of $SO(8,2+n)$ associated to (\ref{geneHO}). For $s=0$, the graded decomposition (\ref{geneHO}) then simplifies to a five-graded decomposition
\begin{multline} \so(8,2+n) \cong \overline{\bf \sfrac{1}{2}(p+2)(p+1)}^\ord{-4} \oplus \scal{ÊÊ \overline{({\bf p+2})} \otimes({\bf 6 +n - 2p}) }^\ord{-2}   \\* \oplus  \scal{Ê\gl_1 \oplus \sl_{2+p} \oplus \so(6-p,n-p)}^\ord{0}  \\* \oplus  \scal{ÊÊ{({\bf p+2})} \otimes({\bf 6 +n - 2p} )  }^\ord{2}  \oplus {\bf \sfrac{1}{2}(p+2)(p+1)}^\ord{4} \end{multline} 
The corresponding orbits of $SO(8,2+n)$ are associated to the $(2+p)$-plets of non-null orthogonal vectors of $SO(6-p,n-p)$ with a given number of time-like vectors $p_+$, and they are commonly labelled as $(+-+)^{p_+}(-+-)^{2+p-p_+}$  \cite{SOpqstrat,SOpqstrat1}.\footnote{There is an extra-degenerance when either the number of time-like vectors in the $(2+p)$-plets is equal to the critical value $6-p$, or the number of space-like vectors to the critical value $n-p$.}  The generic linear combinations of nilpotent elements lying in one of the orbit associated to $(3)^2 (1)^{4+n}$ correspond to linear combinations of $k$ doublet of orthogonal non-null vectors of  $SO(6-p,n-p)$, which define $2k$-plets of non-null vectors.

In order for these linear combinations to give rise to Papapetrou--Majumdar solutions, they must moreover lye in the coset component $\so(8,2+n) \ominus \scal{Ê\so^*(8) \oplus \so(2,n)}$. As a matter of fact, the grade zero component of any graded decomposition of $\so^*(8) \oplus \so(2,n)$ contains at least the compact Lie algebra $\so(4)\oplus \so(n-2)$. It follows from (\ref{geneHO})\footnote{The five-graded decomposition associated to the partition $(2)^{2s}(1)^{10+n-4s}$ being \begin{multline} \so(8,2+n) \cong  \overline{\bf s ( 2s-1)}^\ord{-2}  \oplus   \scal{\overline{\bf 2s} \otimes ({\bf 10 +n  - 4s}) }^\ord{-1}  \oplus \\*   \scal{ \gl_1 \oplus \sl_{2s} \oplus \so(8-2s,2+n-2s)}^\ord{0}    \oplus \scal{Ê{\bf 2s} \otimes ({\bf 10 +n - 4s}) }^\ord{1} \oplus   { \bf s ( 2s-1)}^\ord{2} \nn \end{multline} } that the orbits associated to the partitions   $(3)^{p} (2)^{2s} (1)^{10+ n-3p-4s}$ have no intersection with $\so(8,2+n) \ominus \scal{Ê\so^*(8) \oplus \so(2,n)}$ for $p+2s >4$, and the only higher order nilpotent orbits to consider (with $p+s>2$) are the ones associated to the partitions $(3)^3 (1)^{1+n}$ and $(3)^4 (1)^{n-2}$.

The orbits associated to the partition $(3)^3 (1)^{1+n}$ correspond to triplet of non-null vectors of $SO(5,n-1)$ in the grade two component of the following five graded decomposition of $\so(8,2+n)$,
\be \so(8,2+n) \cong {\bf 3}^\ord{-4} \oplus \scal{Ê  {\bf \bar 3} \otimes   ({\bf 4+n})}^\ord{-2} \oplus  \scal{\gl_1 \oplusÊ\sl_3 \oplus \so(5,n-1)}^\ord{0} \oplus   \scal{Ê  {\bf 3} \otimes   ({\bf 4+n})}^\ord{2} \oplus {\bf \bar 3}^\ord{4} \label{troistrois} \ee
There are four nilpotent orbits associated to this decomposition, each orbit is determined by the number of time-like vectors versus the number of space-like vectors of the triplet. There are only two five graded decompositions of  $Ê\so^*(8) \oplus \so(2,n)$ compatible with this five-graded decomposition of $\so(8,2+n)$, namely 
\begin{multline} \so^*(8) \oplus \so(2,n)  \cong {\bf 1}_{\scriptscriptstyle --}^\ord{-4} \oplus \scal{Ê({\bf 2} \otimes {\bf 4})_- \oplus {\bf n}_{\scriptscriptstyle ++}}^\ord{-2} \oplus \gl_1 \\*\oplus \scal{Ê\so(1,1) \oplus \sl_2 \oplus \so(4) \oplus \so(1,n-1)}^\ord{0} \oplus \scal{Ê({\bf 2} \otimes {\bf 4})_+ \oplus {\bf n}_{\scriptscriptstyle --}}^\ord{2} \oplus  {\bf 1}_{\scriptscriptstyle ++}^\ord{4} \label{Imun} \end{multline}
for which the nilpotent element can be chosen to carry a non-zero component in ${\bf 4}_{\scriptscriptstyle --}^\ord{2}$ and a doublet of orthogonal non-null vectors of $SO(1,n-1)$ inside $({\bf 2} \otimes {\bf n})_+^\ord{2}$; and 
\begin{multline} \so^*(8) \oplus \so(2,n)  \cong {\bf 1}_{\scriptscriptstyle --}^\ord{-4} \oplus \scal{Ê({\bf 2} \otimes ({\bf n-2}))_- \oplus {\bf 6}_{\scriptscriptstyle ++}}^\ord{-2} \oplus \gl_1 \\*\oplus \scal{Ê\so(1,1) \oplus \sl_2 \oplus \so(n-2) \oplus \so(5,1)}^\ord{0} \oplus \scal{Ê({\bf 2} \otimes ({\bf n-2}))_+ \oplus {\bf 6}_{\scriptscriptstyle --}}^\ord{2} \oplus  {\bf 1}_{\scriptscriptstyle ++}^\ord{4}  \label{Imdeux} \end{multline}
for which the nilpotent element can be chosen to carry a non-zero component in $({\bf n-2})_{\scriptscriptstyle --}^\ord{2}$ and a doublet of linearly independent  non-null vectors of $SO(5,1)$ inside $({\bf 2} \otimes {\bf 6})_+^\ord{2}$.

The first embedding (\ref{Imun}) gives rise to two nilpotent $SO(6,2) \times SO(2,n)$ orbits in $\so(8,2+n) \ominus \scal{Ê\so^*(8) \oplus \so(2,n)}$: one $ (+-+)^2 (-+-)$ of isotropy subgroup 
\be \scal{ÊSO(1,1) \times SO(3) \times SO(n-2)} \ltimes \scal{Ê ( {\bf 2} \otimes   {\bf 3} \oplus {\bf 2} \oplus {\bf n-2}  )^\ord{1} \oplus {\bf 1}^\ord{2}} \ee
which interpolates between three nilpotent  $SO(6,2) \times SO(2,n)$ orbits, one $ (+-+)^2 $ and two $ (+-+) (-+-)$ of isotropy subgroup 
\be ISO(4,1) \times ISO(n-1) \times \mathds{R}  \hspace{10mm} \mbox{ and} \hspace{10mm} \begin{array}{c} \Ic\scal{ÊSO(1,1)  \times SO(4)} \times SO(1,n-1)  \\*
 ISO(4,1) \times ISO(1,n-2) \times \mathds{R} \end{array} 
\label{interpolA} \ee
respectively; and one $ (+-+) (-+-)^2$ of isotropy subgroup 
\be \scal{ÊSO(2) \times SO(3) \times SO(1,n-3)} \ltimes \scal{Ê ({\bf 2} \otimes   {\bf 3} \oplus {\bf 2} \oplus {\bf n-2}  )^\ord{1} \oplus {\bf 1}^\ord{2}} \ee
which interpolates between the two nilpotent  $SO(6,2) \times SO(2,n)$ orbits $ (+-+) (-+-) $ and $  (-+-)^2$ of isotropy subgroup 
\be ISO(4,1) \times ISO(1,n-2) \times \mathds{R}   \hspace{10mm} \mbox{ and} \hspace{10mm} \Ic\scal{ÊSO(2) \times SO(4)} \times SO(2,n-2)   \label{interpolC} \ee
respectively. 

The second embedding (\ref{Imdeux}) gives rise to two nilpotent $SO(6,2) \times SO(2,n)$ orbit in $\so(8,2+n) \ominus \scal{Ê\so^*(8) \oplus \so(2,n)}$ as well: one $(+-+)^2 (-+-)$ of isotropy subgroup 
\be \scal{ÊSO(2) \times SO(3,1) \times SO(n-3)} \ltimes \scal{Ê ({\bf 2} \oplus   {\bf 4} \oplus {\bf 2} \otimes ({\bf n-3}) )^\ord{1} \oplus {\bf 1}^\ord{2}} \ee
which interpolates between the two nilpotent  $SO(6,2) \times SO(2,n)$ orbits $ (+-+)^2 $ and $ (+-+) (-+-)$ of isotropy subgroup 
\be SO(4,2) \times \Ic\scal{ÊSO(2) \times SO(n-2)}   \hspace{10mm} \mbox{ and} \hspace{10mm} ISO(4,1) \times ISO(1,n-2) \times \mathds{R}  \label{interpolB}\ee
respectively; and one $ (+-+) (-+-)^2$ of isotropy subgroup 
\be \scal{ÊSO(1,1) \times SO(4) \times SO(n-3)} \ltimes \scal{Ê ({\bf 2} \oplus   {\bf 4} \oplus {\bf 2} \otimes ({\bf n-3}) )^\ord{1} \oplus {\bf 1}^\ord{2}} \ee
which interpolates between three nilpotent  $SO(6,2) \times SO(2,n)$ orbits, two  $ (+-+) (-+-) $ and one $  (-+-)^2$ of isotropy subgroup 
\be \begin{array}{c} SO(5,1) \times \Ic\scal{ÊSO(1,1) \times  SO(n-2)} \\*
ISO(4,1) \times ISO(1,n-2) \times \mathds{R} \end{array} 
  \hspace{10mm} \mbox{ and} \hspace{10mm} ISO(5) \times ISO(1,n-2) \times \mathds{R} \label{interpolD} \ee
respectively. 

As a result, such higher order orbits do not permit to interpolate between charge matrix lying in the $SO(6,2) \times SO(2,n)$ orbits of isotropy subgroup 
\be \begin{array}{c} \Ic\scal{ÊSO(2) \times SO(4)} \times SO(n) \\*
SO(6) \times  \Ic\scal{ÊSO(2) \times SO(n-2)} \end{array}  \hspace{10mm} \mbox{ and} \hspace{10mm} ISO(5) \times ISO(n-1) \times \mathds{R} \ee
which are the ones which correspond to non-singular black holes \cite{nous}. For instance, the appearance of $ISO(4,1)$ in the isotropy group implies that the solutions associated to the corresponding orbit carry one central charge which is larger than the mass, \eg  $p < P \Rightarrow M < Q+P $ in (\ref{NBcharge}); and the appearance of $ISO(1,n-2)$ corresponds in the same way to solutions with a matter electromagnetic charge larger than the mass, \eg $Q< q \Rightarrow M < q+p$ in (\ref{NBcharge}). The isotropy subgroup $ \Ic\scal{ÊSO(1,1)  \times SO(4)} \times SO(1,n-1) $ correspond to BPS solutions for which the $SL(2,\mathds{R}) \times SO(6,n)$ quartic invariant $\scal{Ê|\w|^2 + |z|^2 - z^A \bar z_A }^2 -   \bigl|  2 \w \bar z - z^A z_A \bigr|^2 $ is strictly negative, as the isotropy subgroup $SO(5,1) \times \Ic\scal{ÊSO(1,1) \times  SO(n-2)}$ corresponds to extremal solutions carrying one saturated matter charge $z^A \bar z_A + \sqrt{ ( z^A \bar z_A )^2- |z^A z_A|^2 } = 2 |\w|^2$ and a strictly negative $SL(2,\mathds{R}) \times SO(6,n)$ quartic invariant.

The orbits associated to the partition $(3)^4 (1)^{n-2}$ correspond to quartet of non-null vectors of $SO(4,n-2)$ lying in the grade two component of the five-graded decomposition
\be \so(8,2+n) \cong {\bf 6}^\ord{-4} \oplus \scal{Ê\overline{\bf 4} \otimes ({\bf 2+n})}^\ord{-2}Ê\oplus  \scal{\gl_1 \oplusÊ\sl_4 \oplus \so(4,n-2)}^\ord{0} \oplus \scal{Ê{\bf 4} \otimes ({\bf 2+n})}^\ord{2} \oplus {\bf 6}^\ord{4} \label{troisquatre} \ee
There is only one five-graded decomposition of $\so(6,2) \oplus \so(2,n)$ compatible with this graded decomposition, which is 
\begin{multline}
 \so(6,2) \oplus \so(2,n) \cong \scal{ {\bf 1}_{++} \oplus {\bf 1}_{--}}^\ord{-4} \oplus \scal{Ê{\bf 2} \otimes {\bf 4}_+ \oplus {\bf 2} \otimes (n-2)_-}^\ord{-2} \\* \oplus    \scal{Ê\gl_1 \oplus \so(1,1) \oplus \sl_2 \oplus \sl_2 \oplus \so(4) \oplus \so(n-2) }^\ord{0} \\* \oplus   \scal{Ê{\bf 2} \otimes {\bf 4}_- \oplus {\bf 2} \otimes (n-2)_+}^\ord{2} \oplus \scal{ {\bf 1}_{++} \oplus {\bf 1}_{--}}^\ord{4}  \label{troisquatre} \end{multline} 
As a result, among the six orbits associated to the partition $(3)^4 (1)^{n-2}$, only the $(+-+)^2 (-+-)^2$ 
 one admits a non-trivial intersection with the coset component $\so(8,2+n) \ominus \scal{Ê\so(6,2) \oplus \so(2,n)}$, leading to one single corresponding $SO(2,6) \times SO(2,n)$ orbit. A similar determination of the $SO(6,2) \times SO(2,n)$ orbits $(3)^2 (1)^{4+n}$  admitting a representative in the component of grade two $({\bf 2} \otimes {\bf 4})_+ \oplus ({\bf 2} \otimes ({\bf n-2}))_-$ shows that they all correspond to singular black holes of the same kind as the one appearing in the case of the $(3)^3 (1)^{1+n}$ orbits.

The last step before to conclude is to check that one can not build multi-black holes solutions involving only black holes with vanishing horizon area that would not correspond to linear  combinations of nilpotent elements lying in an orbit associated to the partition $(3)^2 (1)^{4+n}$. Representatives of the orbits associated to the partition $(3)(2)^2(1)^{3+n}$ within the five-graded decompositions (\ref{troisquatre})  involve a vector of either $SO(4)$ or $SO(n-2)$ of the grade two component, as well as a null-vector of $SO(2,2) \cong SL(2,\mathds{R}) \times_{\mathds{Z}_2}  SL(2,\mathds{R})$ of the grade four component. We computed the associated isotropy subgroups to be 
\bea && \scal{ \mathds{R}_+^* \times SO(4,1) \times SO(n-2) }Ê\ltimes \scal{Ê( {\bf 1} \oplus {\bf 5}Ê\oplus ({\bf n-2}) )^\ord{1} \oplus ({\bf n-2})^\ord{2} \oplus {\bf 1}^\ord{2} } \CR
 && \scal{ \mathds{R}_+^* \times SO(4) \times SO(1,n-2) }Ê\ltimes \scal{Ê( {\bf 1} \oplus {\bf 4}Ê\oplus ({\bf n-1}) )^\ord{1} \oplus {\bf 4}^\ord{2} \oplus {\bf 1}^\ord{2} }
 \eea
They correspond to black holes for which the $SL(2,\mathds{R}) \times SO(6,n)$ quartic invariant vanishes and either, one central charge is saturated and one matter charge is larger than the mass (\eg $p=P$ and $q>Q$), or, one matter charge is saturated and one central charge is larger than the mass (\eg $Q=q$ and $P>p$), respectively. Any linear interpolation of elements of this nilpotent orbit which lye in an orbit associated to the partition $(3)^3(1)^{1+n}$ involves such elements as well. Similarly, one finds that the linear combinations of elements of the nilpotent orbits associated to the partition $(3)(1)^{7+n}$ in the higher order orbits always involve elements of $SO(6,2) \times SO(2,n)$ orbits of isotropy subgroup $SO(5,2) \times ISO(1,n-1)$ or $ISO(5,1) \times SO(2,n-1)$. They correspond to singular black holes which carry either saturated matter charges and central charges larger than the mass (\eg $p=P=0$ and $q>Q$), or saturated central charges and matter charges larger than the mass (\eg $Q=q=0$ and $P>p$). Within the graded decomposition (\ref{troisquatre}), the representatives of  elements of the nilpotent orbits associated to the partitions $(2)^4 (1)^{2+n}$ and $(2)^2 (1)^{6+n}$ lye  in the grade four component $( {\bf 2} \otimes {\bf 2})^\ord{4}$, in such a way that any linear combination of such elements lies in a lower order orbit. Similarly within the graded decompositions associated to the partition $(3)^3 (1)^{1+n}$, such elements only involve two null-vectors, such that any linear combination of them turns out to satisfy the cubic characteristic equation (\ref{cubic}).

We have thus proved that all the solutions of Papapetrou--Majumdar type associated to higher order orbits carry naked singularity, and it follows that the multi-black holes solutions discussed in the preceding section define the most general solutions of Papapetrou--Majumdar type within $\N=4$ supergravity coupled to $n$ vector multiplets. 


As we are going to see, the situation is very similar in maximal supergravity, although we have not completed the proof in this case. The nilpotent orbits of $\e_8$ are labelled by their so-called weighted Dynkin diagram. One can always define the $\sl_2$ triplet (\ref{triplet}) associated to a nilpotent orbit such that the element ${\bf H}_{\rm N}$ lies in a chosen Cartan subalgebra. The triplet is then called a normal triplet \cite{coadjoint}. A $\e_8$ weighted Dynkin diagram coordinatises ${\bf H}_{\rm N}$ as a vector of the Cartan subalgebra of $\e_8$ and determines in a unique way the corresponding complex orbit. The real orbits of $\e_{8(8)}$ are in one to one correspondence with the $Spin(16,\mathds{C})$  orbits in the coset $\e_8 \ominus \so(16,\mathds{C})$ through the Kostant--Sekiguchi correspondence. One can always define the $\sl_2$ triplet associated to a nilpotent orbit such that both ${\bf E}_{\rm C} $ and ${\bf F}_{\rm C}Ê$ lie in $\e_8 \ominus \so(16,\mathds{C})$ and such that ${\bf H}_{\rm C}$ lies in a chosen Cartan subalgebra of $\so(16,\mathds{C})$. The triplet is then called a Caley triplet \cite{coadjoint}. An $\so(16,\mathds{C})$ weighted Dynkin diagram coordinatises ${\bf H}_{\rm C}$ as a vector of the Cartan subalgebra of $\so(16,\mathds{C})$ and determines in a unique way the corresponding real orbit. The $\so(16,\mathds{C})$ weighted Dynkin diagrams associated to the various $E_{8(8)}$ orbits lying in a given complex orbit also determine all the graded decompositions of $\so(16,\mathds{C})$ consistent with the corresponding  graded decomposition of $\e_{8}$. The non-trivial intersection with $\e_{8(8)} \ominus \so^*(16)$ correspond to such graded decompositions that define a graded decomposition of $\so^*(16)$ compatible with the one of $\e_{8(8)}$. Since $\e_{8(8)}$ is split, its Cartan subalgebra is the direct sum of eight copies of $\gl_1$, and there is a graded decomposition of $\e_{8(8)}$ associated to each $\e_8$ weighted Dynkin diagram. On the other hand, $\so^*(16)$ is only half split and its Cartan subalgebra is the direct sum of four copies of $\gl_1(\mathds{C})$ such that only the $\so(16,\mathds{C})$ weighted Dynkin diagrams of the form {{\tiny $ {  \vspace{-2mm} \left[ \begin{array}{ccccccccc}  && \mathfrak{0} \hspace{-0.7mm}&&&&&& \vspace{ -1.5mm} \\ \cdot \hspace{-0.5mm}& \mathpzc{a} \hspace{-0.6mm} &\mathpzc{b} \hspace{-0.6mm} &  \mathfrak{0} \hspace{-0.7mm}& \mathpzc{c} \hspace{-0.6mm} & \mathfrak{0}\hspace{-0.7mm}&\mathpzc{d}\hspace{-0.6mm}&\mathfrak{0} \end{array}\right] }$}} with $\mathpzc{a}, \, \mathpzc{b}, \, \mathpzc{c}, \, \mathpzc{d} \in \mathds{N}$, define graded decompositions of $\so^*(16)$.

Let us give an example. The minimal nilpotent orbit of $\e_{8(8)}$ is associated to the $\e_8$ weighted Dynkin diagram \DEVIII00000001 and the $\so(16,\mathds{C})$ weighted Dynkin diagram \DSOXVI00000010. The associated graded decompositions of $\e_{8(8)}$ and $\so^*(16)$, \ie
\begin{gather} 
\e_{8(8)} \cong {\bf 1}^\ord{-2}Ê\oplus {\bf 56}^\ord{-1} \oplus  \scal{ \gl_1 \oplus \e_{7(7)}}^\ord{0} \oplus {\bf 56}^\ord{1} \oplus {\bf 1}^\ord{2} \CR
\so^*(16) \cong \overline{\bf 28}^\ord{-1}Ê\oplus \scal{Ê\gl_1 \oplus \su^*(8)}^\ord{0} \oplus {\bf 28}^\ord{1} 
\end{gather}
are compatible, such that there is an associated  non-trivial $Spin^*(16)$ orbit, which turns out to be homeomorphic to the moduli space of spherically symmetric \ft12 BPS black holes \cite{nous}. Note that the set of zeros of the weighted Dynkin diagram draws the Dynkin diagram of the grade zero component.  Let us consider the higher order orbit for which the representative vanishes at the fourth power in the adjoint representation. The latter is associated to the weighted Dynkin diagrams \DEVIII01000000 and \DSOXVI10001000 of $\e_8$ and $\so(16,\mathds{C})$, respectively. The associated graded decomposition of $\so(16,\mathds{C})$ does not define a graded decomposition of $\so^*(16)$ because $\sl_4 \oplus \sl_4   \hspace{2mm}  / \hspace{-4mm} \subset \so^*(16)$, and the corresponding real orbit does not intersect with the coset component $\e_{8(8)} \ominus \so^*(16)$. 

Exploiting the tables of \cite{coadjoint,E8strat}, one finds that there are two real orbits of $\e_{8(8)}$ of degree six in the adjoint (\ie which representatives satisfy ${\ad_{\bf E}}^6 = 0$) which do not intersect with the coset component, as well as nine higher order orbits of degree five  (which representatives satisfy ${\ad_{\bf E}}^5 = 0$ and ${\bf E}^5 \ne 0$ in the ${\bf 3875}$), from which only seven admit potentially a non-trivial intersection with the coset component. They are, the two real orbits associated to the weighted Dynkin diagram \DEVIII10000001, which unique compatible seven-graded decomposition of $\so^*(16)$ is associated to the weighted Dynkin diagram \DSOXVI00010010, the two real orbits associated to the weighted Dynkin diagram \DEVIII00000100, which unique compatible nine-graded decomposition of $\so^*(16)$ is associated to the weighted Dynkin diagram \DSOXVI01000100, and the three real orbits associated to the weighted Dynkin diagram \DEVIII20000000, which unique compatible graded decomposition of $\so^*(16)$ is associated to the weighted Dynkin diagram \DSOXVI00020000. The latter decompositions are even (\ie all the entries are even integers), and the associated ${\bf H}_{\rm N}$ is just twice the one associated to the $Spin^*(16)$ orbit of spherically symmetric \ft14 BPS black holes. This is very similar to the case of $\N=4$, since the graded decompositions associated to the partition $(3)^4 (1)^{n-2}$ are also even, and the associated ${\bf H}$ is just twice the one associated to the $SO(6,2) \times SO(2,n)$ orbit of spherically symmetric \ft14 BPS black holes with one matter electromagnetic charge saturated (\eg $Q=q$ and $p=P$), which can be seen as \ft14 BPS black holes of maximal supergravity for $n\le 6$ \cite{nous}. The graded decomposition associated to the  weighted Dynkin diagram \DEVIII20000000,
\be \e_{8(8)} \cong {\bf 14}^\ord{-4} \oplus \overline{\bf 64}^\ord{-2} \oplus \scal{Ê\gl_1 \oplus \so(7,7)}^\ord{0} \oplus {\bf 64}^\ord{2} \oplus {\bf 14}^\ord{4} \ee
can indeed be truncated to (\ref{troisquatre}) by considering the embedding $SO(3,3)  \times Spin(4,4) \subset SO(7,7)$. In the same way, the graded decomposition associated to 
the weighted Dynkin diagram \DEVIII00000100,
\begin{multline}  \e_{8(8)} \cong  {\bf 3}^\ord{-4} \oplus {\bf 16}^\ord{-3} \oplus \scal{Ê  {\bf \bar 3} \otimes   {\bf 10}}^\ord{-2} \oplus \scal{Ê{\bf 3} \otimes \overline{\bf 16}}^\ord{-1} \\* \oplus   \scal{Ê\gl_1 \oplus \sl_3 \oplus \so(5,5)}^\ord{0} \oplus  \scal{Ê\overline{\bf 3} \otimes {\bf 16}}^\ord{1}  \oplus   \scal{Ê  {\bf 3} \otimes   {\bf 10}}^\ord{2} \oplus \overline{ \bf 16}^\ord{3} \oplus {\bf \bar 3}^\ord{4}   \end{multline} 
can be truncated to (\ref{troistrois}) by disregarding the components of odd degree. We thus expect these graded decomposition to only possibly define singular Papapetrou--Majumdar solutions as in the case of $\N=4$. The two real orbits associated to the weighted Dynkin diagram \DEVIII10000001 have no equivalent in $\N=4$, nevertheless, the one associated to the weighted Dynkin diagram \DSOXVI11000001 only contains orbits associated to BPS solutions in its boundary, whereas the one associated to the weighted Dynkin diagram \DSOXVI00010010 does not contains the orbit associated  to the generic \ft18 BPS solutions in its boundary \cite{E8strat}. These orbit thus do not permit to define multi-black hole solutions involving both, generic \ft18 BPS black holes and non-BPS extremal black holes.

\vspace{0.5cm}

\noindent
{\bf Acknowledgments}: I am grateful to Hermann Nicolai, Jakob Palmkvist, Boris Pioline and Kelly Stelle
for discussions and comments.


\end{document}